\documentstyle[aps,preprint,psfig]{revtex}
\begin{document}

\title{Memory function approach to the Hall constant in strongly 
       correlated electron systems}

\author{Ekkehard Lange}

\address{Institut f\"ur Theorie der Kondensierten Materie,
Universit\"{a}t Karlsruhe, 76128 Karlsruhe,\\ Germany\\}

\maketitle

\begin{abstract}
The anomalous properties of the Hall constant in the normal state of 
high-$T_c$ superconductors are investigated within the single-band 
Hubbard model. We argue that the Mori theory is the appropriate 
formalism to address the Hall constant, since it aims directly at 
resistivities rather than conductivities. More specifically, the 
frequency dependent Hall constant decomposes into its infinite 
frequency limit and a memory function contribution. As a first step, 
both terms are calculated perturbatively in $U$ and on an infinite 
dimensional lattice, where $U$ is the correlation strength. If we 
allow $U$ to be of the order of twice the bare band width, the memory 
function contribution causes the Hall constant to change sign as a 
function of doping and to decrease as a function of temperature.\\
\\ 
PACS numbers:

\end{abstract}

\newpage
%\maketitle

\normalsize

\section{Introduction}

Since the discovery of high-$T_c$ superconductors ten years ago, the
anomalous properties of their normal state have been the subject of
intensive theoretical work. It is widely believed that a model of 
strongly correlated electrons already captures the basic ingredients 
of the relevant physics. In these models, the correlations are 
represented by a strong local interaction $U$. However, a coherent 
description of {\em all} anomalous properties on the basis of such a 
model is still lacking. The main problem is that exact calculations
are generally feasible only in a small parameter regime and that most 
approximation schemes fail in capturing {\em all} aspects which are 
supposed to be important.

The Hall constant is especially hard to describe. One reason for this 
is that the Hall conductivity contains a three-point 
correlation function after it has been expanded to first order in the
magnetic field. Then, the calculation of vertex corrections is a tough
problem which, to our knowledge, has been attempted only in the case 
of a Fermi liquid and to leading order in the quasiparticle damping 
\cite{kohno}. Moreover, since the frequency dependent Hall constant 
is given as a quotient of conductivities, the limit 
$\omega\rightarrow0$ may be precarious due to resonances like the 
Drude peak. A more technical peculiarity of the Hall effect is due to
the fact that the magnetic field is introduced via a vector potential
which, formally, breaks the symmetry with respect to lattice
translations. But even in the simplest case of a Bloch-Boltzmann 
description, the temperature dependence of the Hall constant may be
difficult to reproduce, because the relaxation time cancels once it 
is assumed to be independent of momentum.

The measurements of the Hall constant in high-$T_c$ materials
\cite{ong1} reveal two major anomalous dependences: on temperature and
on doping. Both cannot be understood within conventional band
theory. For noninteracting tight-binding electrons on a
two-dimensional square lattice, the Hall constant changes sign at half
filling as the Fermi surface changes its shape from electronlike to 
holelike. In contrast, the Hubbard model in the large-$U$ limit
exhibits an additional sign change below half filling which is purely 
due to correlations \cite{SSS,fukuyama}. In addition, in the limit 
$\delta\rightarrow0$, i.e.\ near half filling, the Hall constant 
diverges according to a $1/\delta$ law \cite{SSS,fukuyama}. These 
properties are supposed to account for the doping dependence observed
in, e.g., La$_{2-x}$Sr$_x$CuO$_4$ \cite{takagi,ong2}. As for the 
anomalous temperature dependence of the Hall constant, the most 
striking features are: firstly, a strong decrease which is, in some 
cases, as fast as $1/T$ \cite{ong1}; secondly, the lack of saturation 
above a fraction of the Debye temperature, typically $\sim0.2-0.4
T_D$ \cite{ong3,harris}, in contrast to what is expected in a Fermi 
liquid description with weak electron-phonon coupling \cite{ziman}; 
and thirdly, a quadratic dependence of the Hall angle on temperature 
for not too large dopings \cite{harris,chien,hwang}. In a Fermi
liquid, the temperature dependence arises from an anisotropic
relaxation time \cite{ong3,ziman}. If we assume scattering off phonons
to be the main inelastic process, a temperature dependence is
conceivable only below a certain temperature scale: Then, a
sufficiently anisotropic Fermi surface causes the scattering to be
confined to those regions of the Fermi surface, where small momentum 
transfers are possible. For high enough temperatures, this kinematic 
restriction is lifted and the scattering becomes isotropic, thus
leading to a cancellation of the relaxation time. The cross-over 
temperature is given by $\sim0.2-0.4 T_D$. The universally observed 
decrease of the Hall constant as a function of temperature in almost 
all high-$T_c$ compounds up to temperatures clearly beyond this 
temperature scale must therefore be due to electronic correlations as 
well.

In the following, we investigate the Hall effect on the basis 
of the simplest model of strongly correlated electrons, namely the 
single-band Hubbard model on a hypercubic lattice in $d$ dimensions
with nearest-neighbour hopping. This model, along with Mori's formalism 
used to represent the Hall constant, is introduced in Sec.\ II. In this
theory, the Hall constant is given as the sum of its infinite
frequency limit ($R_H^{\infty}$) and a memory function contribution. 
The former term was first considered by Shastry, Shraiman and Singh 
\cite{SSS}. Our emphasis is on the memory function term which
represents the deviation of the Hall constant from $R_H^{\infty}$ for 
finite frequencies and thus cannot be neglected when considering the
case of zero frequency. The advantage of our representation of the
Hall constant is that we do not have to cope with a quotient of 
conductivities as opposed to the usual approaches. This is why the
Hall constant at low frequencies becomes less sensitive to the
detailed resonance structure of the conductivities. In the remainder
of this paper, this advantage is exploited for the range of weak to 
intermediate correlation strengths, while the opposite limit of strong
correlations will be addressed in a forthcoming paper \cite{lange}.
In Sec.\ III, we proceed by calculating the memory function to second 
order in the Hubbard interaction and to first order in the magnetic 
field. Expansion with respect to the magnetic field leads to a 
decomposition of the memory function into two terms, namely a
two-point and a three-point correlation function. Both contributions 
are evaluated exactly in infinite spatial dimensions. Our results 
indicate that the memory function term is important. Only
then, a precursor effect of the sign change of the Hall constant 
as a function of doping appears even in perturbation theory. 
Moreover, when extrapolating our results to $U$ values of the order 
of twice the free band width $W$, we get most of the qualitative
features observed in, e.g., La$_{2-x}$Sr$_x$CuO$_4$: the sign change 
with respect to doping and the decrease of the Hall constant up to 
unusually high temperatures, characteristic of most high-$T_c$ 
compounds. Finally, in Sec.\ IV, we summarize our main results.

\section{Theoretical framework}
 
\subsection{Single-band Hubbard model}

The single-band Hubbard model on a d-dimensional hypercubic lattice 
in a magnetic field reads
\begin{eqnarray}
   \hat{H}&=&\hat{T}+\hat{V}\,,
  \label{hubbard}\\
   \hat{T}&=&-t\sum_{<ij>} P_{ij}c_{i\sigma}^+c_{j\sigma}\,,
  \label{hopping}\\
   \hat{V}&=&U\sum_i \hat{n}_{i\uparrow}\hat{n}_{i\downarrow}\,,
  \label{interaction}
\end{eqnarray}
where the sum in the hopping term $\hat{T}$ is restricted to nearest 
neighbours and $\hat{V}$ is the Hubbard repulsion. The Peierls phase 
factor $P_{ij}=\exp(ie\int_j^i\,\vec{A}(t,\vec{r}\,)\;d\vec{r}\,)$ 
guarantees the gauge invariance \cite{peierls} 
and the sign of the charge $e$ is chosen to be negative. Since only 
nearest neighbour hops are taken into account, we may approximate
\begin{equation}
 P_{ij}\simeq e^{ie\vec{A}(\vec{R}_j)(\vec{R}_i-\vec{R}_j)}\;,
\label{peierlsapprox}
\end{equation}
where $R_i$ denotes the lattice vector to site $i$. The vector
potential decomposes into two terms describing the electric and
magnetic field, respectively:
\begin{eqnarray}
 \vec{A}(t,\vec{r}\,)&=&\vec{A}^{el}(t)+\vec{A}^{mag}(\vec{r}\,)\,,\\
 \vec{E}(t)&=&-\frac{\partial}{\partial t}\vec{A}^{el}(t)\,,\\
 \vec{H}&=&rot\,\vec{A}^{mag}(\vec{r}\,)\,.
\end{eqnarray}
In linear response theory with respect to the electric field, the
latter appears only in the definition of the current operator. More 
precisely, the current operator is defined as the following functional
derivative:
\begin{equation}
 \hat{J}_{\nu}:=-\frac{1}{e}\frac{\delta\hat{H}(t)}
   {\delta A_{\nu}^{el}(t)}\Bigg|_{A^{el}=0\,.}
\end{equation}
The homogeneous magnetic field is chosen to point in the $z$-direction and 
it is advantageous to fix the gauge from the very beginning according
to the Landau choice
\begin{equation}
 \vec{A}^{mag}(\vec{R})=R_xH\hat{y}\,,
\label{gauge}
\end{equation}
since then, translational symmetry is broken only in one dimension,
namely the $x$-direction. $\hat{y}$ is a primitive lattice vector. 
We need the current operator only up to first order in the magnetic 
field:
\begin{eqnarray}
 \hat{J}_{\nu}&=&\hat{J}_{\nu}^{(0)}+\delta\hat{J}_{\nu}\,,
\label{current}
\\ 
 \hat{J}_{\nu}^{(0)}&=&it\sum_{\vec{R},\vec{R}+\vec{\delta},\sigma}\,
                     \delta_{\nu}\,c_{\sigma}^+(\vec{R}+\vec{\delta})
                                   c_{\sigma}(\vec{R}),
\label{freecurrent}
\\
 \delta\hat{J}_{\nu}&=&-et\sum_{\vec{R},\vec{R}+\vec{\delta},\sigma}\,
                     \delta_{\nu}\,\vec{\delta}\vec{A}^{mag}(\vec{R})
                                 c_{\sigma}^+(\vec{R}+\vec{\delta})
                                 c_{\sigma}(\vec{R})\;.
\label{perturbedcurrent}
\end{eqnarray}
Here, $\vec{\delta}$ is a nearest neighbour vector and the summation is
over pairs of nearest neighbours. Note, however, that due to the gauge
fixation (\ref{gauge}), we cannot choose periodic boundary conditions
in the $x$-direction. Thus, if $\vec{\delta}$ points in the
$x$-direction, we have to carry out the sums in such a way that the 
components $R_x$ and $R_x+\delta_x$ are simultaneously elements of the
set consisting of the $x$-coordinates of all lattice sites, e.g.\ 
$\{R_x^{min},R_x^{min}+1,\ldots,R_x^{max}\}$. Here, it is assumed that
the lattice has $N_x$ sites in the $x$-direction which implies
$R_x^{max} \equiv R_x^{min}+N_x-1$. Of course, observable quantities
are not allowed to depend on the lattice location $R_x^{min}$. The 
hopping term is expanded analogously, yielding:
\begin{eqnarray}
 \hat{T}&=&\hat{T}^{(0)}+\delta\hat{T}\,,
\label{hopping}\\ 
 \hat{T}^{(0)}&=&-t\sum_{\vec{R},\vec{R}+\vec{\delta},\sigma}\,
                                 c_{\sigma}^+(\vec{R}+\vec{\delta})
                                 c_{\sigma}(\vec{R}),
\label{freehopping}\\
 \delta\hat{T}&=&-iet\sum_{\vec{R},\vec{R}+\vec{\delta},\sigma}\,
                                 \vec{\delta}\vec{A}^{mag}(\vec{R})
                                 c_{\sigma}^+(\vec{R}+\vec{\delta})
                                 c_{\sigma}(\vec{R})\;.
\label{perturbation}
\end{eqnarray}
The term without magnetic field, i.e.\ Eq.\ (\ref{freehopping}),
becomes diagonal in crystal momentum space with a band dispersion
$\epsilon_{\vec{k}}=-2t\sum_{j=1}^d \cos k_j$.

\subsection{Mori theory}

In this subsection, the basics of Mori's memory function formalism is
reviewed briefly. For further details, see e.g.\ Ref.\ \cite{fick}.
The best known application of Mori theory is the description of many
particle systems in the hydrodynamic regime \cite{forster}. There, one
is only interested in the dynamics of the hydrodynamic
variables. They are characterized by the fact that their transport is
restricted by conservation laws or by broken symmetries. Thus, they
are bound to vary on a time scale that is very slow in comparison to
that of all the other degrees of freedom. Now, the Mori theory enables
one to separate these two time scales: The equations of motion of the
hydrodynamic variables take on the form of coupled
integro-differential equations. The corresponding integral kernels 
of these so called Mori equations are memory functions in which the 
influence of all the other degrees of freedom is accumulated, hence
the name ``memory function''. In this context of hydrodynamics, the 
memory functions are rapidly varying functions, whose effect may be 
simulated by damping constants. Then, the Mori equations take on a
form analogous to that of the Langevin equation for a particle 
undergoing Brownian motion. However, the validity of the Mori
equations is not restricted to the special set of hydrodynamic
variables. In the simplest case, one sets up the Mori theory for those
observables that constitute the correlation functions one is
interested in. This leads to representations for the unknown
correlation functions in terms of memory functions in which all
analytic properties are fulfilled by construction. On the other hand,
it may be difficult to find an approximate expression for a given
memory function. 

\subsubsection{basic notions}

The {\em Liouville space} ${\cal L}$ is defined as the linear vector 
space over the field of complex numbers whose elements are the linear 
operators in the familiar Hilbert space of quantum mechanics, and 
where the usual operations like scalar multiplication etc. hold. In 
this Liouville space exist linear operators that are called 
{\em superoperators} to distinguish them from the usual ones. 
(Henceforth, normal operators are denoted with a hat, superoperators 
not.) The most important superoperator is the {\em Liouville operator}
$L$, which maps a given operator onto its commutator with the
Hamiltonian: 
\begin{equation}
 L\hat{A}=[\hat{H},\hat{A}]\,.
\end{equation}
An other important class of superoperators are {\em superprojectors}.
However, their definition implies a scalar product in ${\cal L}$. In
the context of response functions, the most convenient scalar product
turns out to be the so called {\em Mori product}
\begin{equation}
 (\hat{A}|\hat{B}):= \frac{1}{\beta}\int_0^{\beta}\,d\tau\,
                     <\;e^{\tau L}\hat{A}^+\cdot\hat{B}\;>\;,
 \label{moriprod}
\end{equation}
where $<\ldots>$ denotes the thermal average and $\beta$ is the
inverse temperature. On the basis of this scalar product, we may now 
speak of adjoint superoperators $S$ and $S^+$, and thus of unitary and
Hermitian ones in the usual sense. The projector $P$ that projects
onto the subspace of ${\cal L}$ spanned by linearly independent
elements $|\hat{G}_i)$, reads:
\begin{equation}
 P=\sum_{ij}|\hat{G}_i)g_{ij}(\hat{G}_j|,
\end{equation}
where the metric $g_{ij}$ is the inverse of the matrix 
$(\hat{G}_i|\hat{G}_j)$, i.e.\ $\sum_k\,g_{ik}
(\hat{G}_k|\hat{G}_j)=\delta_{ij}$. In fact, this implies the
idempotence property $P^2=P$. Finally, the definition of the Mori
product implies the validity of the so called Kubo identity
\begin{equation}
 \beta(\hat{A}|L|\hat{B})=<[\hat{A}^+,\hat{B}]>\;,
\label{kubo}
\end{equation}
which will play an important role.

\subsubsection{Memory function approach to the Hall constant}

From Eq.\ (\ref{kubo}) follows a representation for the 
current-current correlation function $\ll\hat{J}_{\nu}\,;
\,\hat{J}_{\mu}\gg_z$\,, defined as the Laplace transform of
$-i<[e^{iLt}\hat{J}_{\nu}\,,\,\hat{J}_{\mu}]>$:
\begin{equation}
 \ll\hat{J}_{\nu}\,;\,\hat{J}_{\mu}\gg_z=-\beta(\hat{J}_{\mu}|
         \frac{L}{z+L}|\hat{J}_{\nu})\equiv-\chi_{\mu\nu}(z)\,.
\label{currcorr}
\end{equation}
Here, $z$ is a complex frequency, which ultimately has to be 
specialized to $\omega+i0^+$. For formal manipulations, however, it is
more convenient to deal with the complex frequency $z$ rather than 
with $\omega$. The last expression has to be inserted into the Kubo 
formula for the conductivity tensor, 
\begin{equation}
 \sigma_{\nu\mu}(z)=\frac{ie^2}{Nz}\{<\hat{\tau}_{\nu\mu}>
                +\ll\hat{J}_{\nu}\,;\,\hat{J}_{\mu}\gg_z\}\,,
\label{kuboformula}
\end{equation}
where $<\hat{\tau}_{\nu\mu}>$ arises from the equilibrium part of the
current and is defined as the second functional derivative of the
Hamiltonian with respect to the external electric field: 
\begin{equation}
 \hat{\tau}_{\nu\mu}:=\frac{1}{e^2}\frac{\delta^2\hat{H}(t)}
 {\delta A_{\nu}^{el}(t)\,\delta A_{\mu}^{el}(t)}\Bigg|_{A^{el}=0\;.}
\end{equation}
In the Hubbard model (\ref{hubbard}), its expectation value is given
as $<\hat{\tau}_{\nu\mu}>=
2t\sum_{\vec{k}\sigma}cos\,k_x<\hat{n}_{\vec{k}\sigma}>$, 
i.e.\ as the average kinetic energy per dimension.
Using the fact $\lim_{z\rightarrow0}z\,\sigma_{\nu\mu}(z)=0$ which
holds for a metal in the normal state, we show that 
$<\hat{\tau}_{\nu\mu}>$ also equals the static susceptibility
$\chi^0$, defined through 
$\chi_{\mu\nu}(z=0)\equiv\delta_{\mu\nu}\,\chi^0\,$:
\begin{equation}
 \chi^0=\beta(\hat{J}_x|\hat{J}_x)
       =<\hat{\tau}_{xx}>
       =-\frac{1}{d}<\hat{T}>\;.
\label{staticsus}
\end{equation}
Thus, the conductivity tensor may be written as
\begin{equation}
 \sigma_{\nu\mu}(z)=\frac{e^2}{N}\beta(\hat{J}_{\mu}|\frac{i}{z+L}|
               \hat{J}_{\nu})\equiv\frac{e^2}{N}C_{\mu\nu}(z)\,.
\label{cond}
\end{equation}
In order to represent the relaxation functions $C_{\mu\nu}(z)$ in
terms of memory functions, we introduce the superprojector $P$ that 
projects onto the subspace of ${\cal L}$ spanned by the current 
operators $\hat{J}_x$ and $\hat{J}_y$:
\begin{equation}
 P=\frac{\beta}{\chi^0}\sum_{\nu=x,y}|\hat{J}_{\nu})(\hat{J}_{\nu}|\,,
\end{equation}
and the complementary superprojector $Q=1-P$. By making use of the
operator identity $\frac{1}{a+b}=\frac{1}{a}-\frac{1}{a}b
\frac{1}{a+b}$ with $a\equiv z+LQ$ and $b\equiv LP$, we find: 
\begin{eqnarray}
 C_{\mu\nu}(z)&=&\frac{i}{z}\chi^0\delta_{\mu\nu}-\frac{1}{z}
                R_{\mu\alpha}(z)C_{\alpha\nu}(z)\,,\label{relax1}\\
 R_{\mu\nu}(z)&\equiv&\frac{\beta}{\chi^0}(\hat{J}_{\mu}|
                       \frac{z}{z+LQ}L|\hat{J}_{\nu})\\
              &\equiv&\Omega_{\mu\nu}+iM_{\mu\nu}(z)\,.
\end{eqnarray}
The terms in the last equation are the frequency and the memory matrix,
respectively:
\begin{eqnarray}
 \Omega_{\mu\nu}&\equiv&\frac{1}{\chi^0}<[\hat{J}_{\mu}\,,\,
                                   \hat{J}_{\nu}]>\;,
\label{frequency}\\
 M_{\mu\nu}(z)&\equiv&\frac{\beta}{\chi^0}(QL\hat{J}_{\mu}|
                    \frac{i}{z+QLQ}|QL\hat{J}_{\nu})
\label{memory}\,.
\end{eqnarray}
The last equation shows that the dynamics of the memory functions is
governed by $QLQ$ rather than $L$. Solving Eq.\ (\ref{relax1}) for the 
matrix ${\bf C}(z)$ leads to
\begin{equation}
 {\bf C}(z)=i\chi^0\left(z{\bf 1}+{\bf \Omega}+i{\bf M}(z)\right)^{-1}\,. 
\label{relax2}
\end{equation}
Together with Eq.\ (\ref{cond}), this demonstrates that the Mori
theory enables us to calculate directly the resistivity
tensor. Therefore, the desired representation for the dynamical Hall 
constant can be read off from the last equation:
\begin{equation}
 R_H(z)=\frac{N}{ie^2\chi^0}\lim_{H\rightarrow0}\frac{
        \Omega_{xy}+iM_{xy}(z)}{H}\;.
\label{hall}
\end{equation}
Since the memory function term will be shown to vanish in the high
frequency limit as $1/z^2$, the first term represents the high
frequency limit of the Hall constant considered by Shastry et
al. \cite{SSS}:
\begin{equation}
 R_H^{\infty}=\frac{N}{ie^2\chi^0}\lim_{H\rightarrow0}\frac{
        \Omega_{xy}}{H}\;.
\label{R_Hinfty}
\end{equation}
Moreover, $\Omega_{xy}$ is the generalization of the cyclotron 
frequency to the lattice case. Therefore, within a Boltzmann 
equation approach, only the term (\ref{R_Hinfty}) is considered. 
The goal of the subsequent sections is to investigate the memory 
function term $M(z)\equiv M_{xy}(z)$ for finite frequencies.

\subsubsection{Analytic properties}

The analytic properties of $M(z)$ in Eq.\ (\ref{hall}) may all 
be derived on the basis of Eq.\ (\ref{memory}). An alternative 
procedure is to solve Eq.\ (\ref{relax2}) for ${\bf M}(z)$ and go  
back to the analytic properties of the current susceptibilities 
$\chi_{\mu\nu}(z)$, cf.\ Eq.\ (\ref{currcorr}). $M(z)$ reads in 
terms of the susceptibilities $\chi_{\mu\nu}(z)$:
\begin{equation}
 iM(z)=\frac{z\chi^0\chi_{xy}(z)}{(\chi^0-\chi_{xx}(z))^2}-
            \frac{<[\hat{J}_x\,,\,\hat{J}_y]>}{\chi^0}\;.
\label{memsus}
\end{equation} 
From time reversal invariance, homogenity of time and the fact that
the current operators are Hermitian, we may deduce the following 
symmetry properties \cite{forster}:
\begin{eqnarray}
 \chi_{xx}(-z)&=&\chi_{xx}(z)\,,\\
 \chi_{xx}^*(z)&=&\chi_{xx}(z^*)\,,\\
 \chi_{xy}(-z)&=&-\chi_{xy}(z)\,,\label{symxyminus}\\
 \chi_{xy}^*(z)&=&-\chi_{xy}(z^*)\label{symxyconju}\,.
\end{eqnarray} 
Together with Eq.\ (\ref{memsus}), this implies:
\begin{eqnarray}
 M(-z)&=&M(z)\,,\\
 M^*(z)&=&M(z^*)\,.
\label{memprop1}
\end{eqnarray}
$M(z)$ can be represented as a spectral integral
\begin{equation}
 M(z)=\int\frac{d\omega}{\pi}\,\frac{M''(\omega)}{\omega-z}\,,
\end{equation} 
where the spectral function $M''(\omega)$ is given by the
discontinuity across the real axis:
\begin{equation}
 M(\omega\pm i0^+)=M'(\omega)\pm iM''(\omega)\,.
\label{spectral}
\end{equation} 
From the analytic properties (\ref{memprop1}), it follows that 
$M'(\omega)$ and $M''(\omega)$ are real functions satisfying
\begin{eqnarray}
 M'(-\omega)&=&M'(\omega)\,,\\
 M''(-\omega)&=&-M''(\omega)\,.
\end{eqnarray}
Thus, two further conclusions can be drawn: Firstly, only 
even powers in $1/z$ contribute to the high frequency expansion 
of $M(z)$. And secondly, the quotient $M''(\omega)/\omega$ must be
integrable around $\omega=0$,
\begin{equation}
 \int_{-\infty}^{\infty}d\omega\,M''(\omega)/\omega<\infty,
\end{equation}
which can be seen from the fact that the dc-Hall constant contains 
this integral. Note, that we need not understand this expression as  
a Principal value integral due to the fact that the integrand is 
even.

\section{Perturbation theory}

Despite many interesting works on the normal state Hall effect of 
high-$T_c$ superconductors, a calculation that incorporates all the
complicated many-body correlations exactly within a microscopic 
model is still lacking. The following treatment of the Hall constant 
closes this gap at least in the perturbation-theoretical regime. 
On the other hand, the relevant parameter regime is believed to be 
the strong correlation limit rather than the weak one. 
However, it turns out that the final 
expression may well describe the observed dependencies on temperature 
and on doping at least qualitatively, if we allow $U$ to be
extrapolated to values of the order of the free band width $W$.
Thus, a precursor effect of the anomalous dependences clearly shows 
up even in the regime of weak correlations.

\subsection{Approximation}

The perturbation-theoretical treatment of the Hall constant is by no  
means straightforward. As is well known, the evaluation of response 
functions like $\chi_{\mu\nu}(z)$ of Eq.\ (\ref{currcorr}) by
expansion in a small interaction parameter fails because, as an
artefact of such an expansion, these functions become singular for
small frequencies $z$. This difficulty was resolved by G\"otze and 
W\"olfle \cite{woelfle} some time ago by means of a memory function 
approach. They calculated the memory function perturbatively, which, 
at first, is valid only at high enough frequencies. It turns out,
however, that their expression for the memory function depends only 
smoothly on frequency below a certain frequency scale and tends to a 
constant in the limit $\omega\rightarrow0$. Furthermore, in their 
approximation scheme the correct resonance structure of the studied 
response functions is inherently built in. Thus, their results could 
be used in the whole frequency regime including the hydrodynamic one.  
However, we cannot carry over their analysis straightforwardly to the
present problem, because otherwise, we would encounter a spurious 
singularity in the limit that is of most interest, i.e.\ 
$\omega\rightarrow0$.
 
In this subsection, we identify the precondition which is necessary  
to obtain regular expressions in this limit and that was fulfilled 
trivially in the applications of Ref.\ \cite{woelfle} 
but is not in our case. Since this condition does not affect the
correct description of the local interaction $U$, we may take it
as an approximation. By using Mori's formalism, we shall see that 
once this condition is assumed to be satisfied no further
approximations 
have to be made. This last point cannot be seen in the more intuitive 
introduction of the memory function concept as given in Ref.\ 
\cite{woelfle} and shows that the extrapolation to low frequencies 
therein is exact.\\

Perturbation theory is based on the following decomposition of the 
Liouville operator:
\begin{equation}
 L=L_0+L_1\,,
\label{decomposition}
\end{equation}
where $L_0$ and $L_1$ are assigned to the hopping and interaction
term of the Hubbard hamiltonian (\ref{hubbard}), respectively. The
perturbation-theoretical regime is given by the condition $U\ll W$, 
where $W$ is the width of the free band and thus represents the
characteristic energy scale introduced by $L_0$. The precondition to
obtain a regular expression for the memory function $M(z)$ for all 
frequencies to leading order in $U$ is that the relevant operators 
$\hat{J}_x$ and $\hat{J}_y$ span a subspace of ${\cal L}$ that is 
invariant with respect to actions of $L_0$ \cite{fick}. If this
condition were satisfied in our case, it would take on the form
\begin{eqnarray}
 L_0\hat{J}_{\nu}=[\hat{T},\hat{J}_{\nu}]=\hat{J}_{\mu}^{(0)}\;
                                       \Omega_{\mu\nu}^0\,,
\label{appr}
\end{eqnarray}
where $\nu,\mu=x,y$ and summation over repeated indices is implied. 
This can be checked by inserting these equations into
$(\hat{J}_{\lambda}|\ldots)_0$ and comparing the result with the
definition (\ref{frequency}). $(\ldots|\ldots)_0$ is the Mori 
product with respect to $L_0$. Henceforth, bracketed indices refer 
to the magnetic field and unbracketed ones to the decomposition 
(\ref{decomposition}). Unfortunately, the conditions (\ref{appr}) are not
satisfied in the Hubbard model. Instead, we derive with the help of
Eqs.\ (\ref{current})-(\ref{perturbedcurrent}) and
(\ref{hopping})-(\ref{perturbation}) (see the appendix):
\begin{equation} 
 [\hat{T},\hat{J}_x]\,=\frac{\sum_{\vec{k}\sigma}\,
                    \cos k_x\,\sin k_y\,\hat{n}_{\vec{k}\sigma}}
                                {\sum_{\vec{k}\sigma}\,
                    \cos k_x\,\cos k_y\,n_{\vec{k}\sigma}^0}\;
                                <[\hat{J}_y,\hat{J}_x]>_0\;,
\label{condition1}
\end{equation} 
which should be equal to 
\begin{equation} 
 \frac{\sum_{\vec{k}\sigma}\,\sin k_y\,\hat{n}_{\vec{k}\sigma}}
                                {\sum_{\vec{k}\sigma}\,
                             \cos k_y\,n_{\vec{k}\sigma}^0}\;
                                <[\hat{J}_y,\hat{J}_x]>_0
  =\hat{J}_y^{(0)}\Omega_{yx}^0\;.
\end{equation}
This is obviously not the case. Similarly, we find 
\begin{equation} 
 [\hat{T},\hat{J}_y]\,=\frac{\sum_{\vec{k}\sigma}\,
                    \sin k_x\,\cos k_y\,\hat{n}_{\vec{k}\sigma}}
                                {\sum_{\vec{k}\sigma}\,
                    \cos k_x\,\cos k_y\,n_{\vec{k}\sigma}^0}\;
                                <[\hat{J}_x,\hat{J}_y]>_0\;,
\label{condition2}
\end{equation} 
instead of
\begin{equation} 
 \frac{\sum_{\vec{k}\sigma}\,\sin k_x\,\hat{n}_{\vec{k}\sigma}}
                                {\sum_{\vec{k}\sigma}\,
                             \cos k_x\,n_{\vec{k}\sigma}^0}\;
                                <[\hat{J}_x,\hat{J}_y]>_0
  =\hat{J}_x^{(0)}\Omega_{xy}^0\;.
\end{equation}
However, the conditions (\ref{appr}) become exact in the continuum
limit or in the limit of small band fillings. This is seen if we
take explicitly into account the lattice spacing $a$ in the 
arguments of the trigonometric functions which we tacitly have set 
equal to 1. Then we may expand $\cos k_{\nu}a\simeq1$ and $\sin 
k_{\nu}a\simeq k_{\nu}a$ which proves the statement immediately.
Thus, the violation of the conditions (\ref{appr}) on the lattice 
reflects its reduced symmetry in comparison with free space. 
Since the conditions (\ref{appr}) are properties of the {\em free} 
model, we may assume their approximate validity without taking the 
risk of not describing the local interactions (\ref{interaction}) 
correctly.\\

Before proceeding, we show how the conditions (\ref{appr}) appear
within the formalism outlined in Ref.\ \cite{woelfle}. We expand 
Eq.\ (\ref{memsus}) in the frequency regime, where the expression 
$|\chi_{xx}(z)/\chi^0|$ is very small, i.e.\ for high enough 
frequencies and use a couple of times equations of motion for 
correlation functions $\ll\hat{A};\hat{B}\gg_z$. Thus we may show 
that $M(z)$ can be represented as follows (cf.\ Eq.\ (\ref{KKK}):
\begin{equation}
 i\chi^0M(z)\simeq-\frac{\ll\hat{K}_x;\hat{K}_y\gg_z^0}{z}+R(z)\,.
\end{equation}
The first term will be investigated in the next subsection 
and turns out to be regular for all frequencies; $R(z)$ can 
be written as:
\begin{eqnarray}
 R(z)&=&\ll\hat{J}_x;[\hat{T},\hat{J}_y]\gg_z-
        \ll[\hat{T},\hat{J}_x];\hat{J}_y\gg_z
        +2\Omega^0_{xy}\chi_{xx}(z)\nonumber\\
     &=&\frac{2<[\hat{J}_x,\hat{J}_y]>_0}
         {\chi^0}\frac{\phi_{xx}(z)-\phi_{xx}(0)}{z^2}\;,
\end{eqnarray} 
where $\phi_{xx}(z)\equiv\ll\hat{K}_x;\hat{K}_x\gg_z^0$. Calculating
the function $\phi_{xx}(z)$ following the lines outlined in the next 
subsection, we may prove that $R(z)$ is indeed divergent in the limit 
$z\rightarrow0$, which, however, is an artefact of perturbation
theory. The first representation of $R(z)$ in the last set of
equations shows that the condition (\ref{appr}) implies $R(z)$ to 
vanish identically, if we take into account the symmetry properties 
$\Omega_{xy}^0=-\Omega_{yx}^0$ and $\chi_{xx}(z)=\chi_{yy}(z)$.

\subsection{Reduction to ordinary correlation functions}

We are interested in the memory function appearing in
Eq.\ (\ref{hall}), whose Laplace transform is given according to
Eq.\ (\ref{memory}) as
\begin{equation}
 M(t)=\frac{\beta}{\chi^0}(QL\hat{J}_x|e^{iQLQt}|QL\hat{J}_y)\,.
\end{equation}
Due to the approximation (\ref{appr}), the free part $L_0$ of the
Liouville operator does not contribute to the operator
$QL|\hat{J}_{\nu})$. Hence, to leading order in the interaction
strength $U$, we obtain
\begin{equation}
 M(t)=\frac{\beta}{\chi^0}(Q_0L_1\hat{J}_x|e^{iQ_0L_0Q_0t}
                                          |Q_0L_1\hat{J}_y)_{0\,.}
\end{equation}
Since $Q_0$ commutes with $L_0$ and because of the idempotence of 
$Q_0$ \cite{fick}, we may free ourselves of all superprojectors $Q_0$ with
the exception of one, say, that within the ``ket'' $|Q_0L_1\hat{J}_y)$.
However, even this last appearance of $Q_0$ may be omitted, since its
part $P_0$ leads to a term proportional to the following first order 
expression of the frequency matrix \cite{fick}: $\Omega_{\nu\mu}^1=
g^0_{\nu\lambda}(\hat{J}_{\lambda}|L_1|\hat{J}_{\mu})_0$. Here 
summation over equal indices is implied. However, the frequency 
matrix is easily traced back to $n_{\vec{k}\sigma}=
<c_{\vec{k}\sigma}^+c_{\vec{k}\sigma}>$, whose first order
contribution vanishes. Thus, $\Omega_{\nu\mu}^1$ vanishes as stated
and we arrive at
$M(z)=(\beta/\chi^0)(\hat{K}_x|i/(z+L_0)|\hat{K}_y)_0$, where
we have defined
\begin{equation}
 \hat{K}_{\nu}\equiv[\hat{V},\hat{J}_{\nu}]\,.
\label{KKK}
\end{equation}
With the identity $z/(z+L_0)=1-L_0/(z+L_0)$ and the symmetry property
$\ll\hat{K}_x;\hat{K}_y\gg_{-z}^0=-\ll\hat{K}_x;\hat{K}_y\gg_z^0$,
which may be traced back to Eq.\ (\ref{symxyminus}) by means of two 
equations of motion, we eventually arrive at
\begin{equation}
 i\chi^0M(z)=\,-\;\frac{\ll\hat{K}_x;\hat{K}_y\gg_z^0}{z}\;.
\label{memory1}
\end{equation}
Now, we must evaluate this correlation function for the free tight
binding model (\ref{hopping}) to leading first order in the magnetic  
field. In order to
derive explicit expressions for the operators (\ref{KKK}), we
introduce the following combination of Blochoperators: 
\begin{equation}
 \hat{A}^{\sigma}_{\vec{k}_1,\vec{k}_2|\vec{k}|\vec{q}}\equiv
    c_{\vec{k}_1\sigma}^+c_{\vec{k}_2\sigma}
    c_{\vec{k}-\vec{q}|-\sigma}^+c_{\vec{k}|-\sigma}\;.
\label{basicblock}
\end{equation}
This is the basic building block for the operators $\hat{K}_{\nu}$. To
see this, we insert Eqs.\ (\ref{interaction}) and
(\ref{current})-(\ref{perturbedcurrent}) into the definition
(\ref{KKK}) and write the result in terms of Blochoperators. We find:
\begin{eqnarray}
 \hat{K}_{\nu}&=&\hat{K}_{\nu}^0+\delta\hat{K}_{\nu}\,,\\
 \hat{K}_{\nu}^{(0)}&=&-\frac{U}{N}\sum_{\vec{k}_1\vec{k}_2\vec{k}\vec{q}
  \sigma}\;\left[B_{\nu}(\vec{k}_1,\vec{k}_2+\vec{q}\,)
           -B_{\nu}(\vec{k}_1-\vec{q},\vec{k}_2)\right]
            \hat{A}^{\sigma}_{\vec{k}_1,\vec{k}_2|\vec{k}|\vec{q}\;,}
\label{freeK}\\
 \delta\hat{K}_x&=&0\;,\\
 \delta\hat{K}_y&=&Het\frac{U}{N}\sum_{\vec{k}_1\vec{k}_2\vec{k}\vec{q}
  \sigma}\;\left[C(\vec{k}_1,\vec{k}_2+\vec{q}\,)
           -C(\vec{k}_1-\vec{q},\vec{k}_2)\right]
            \hat{A}^{\sigma}_{\vec{k}_1,\vec{k}_2|\vec{k}|\vec{q}\;,}
\label{perturbedK}
\end{eqnarray}
where the matrices $B_{\nu}(\vec{k}_1,\vec{k}_2)$ and
$C(\vec{k}_1,\vec{k}_2)$ are defined in the appendix, cf.\ Eqs.\ 
(\ref{Bsubnu}) and (\ref{CCC}). Since $\delta\hat{K}_x$ vanishes, the 
expansion of the correlation function of Eq.\ (\ref{memory1}) to first
order in the magnetic field reads:
\begin{eqnarray}
 \ll\hat{K}_x;\hat{K}_y\gg_z^0&=&C^{II}(z)+C^{III}(z)\,,
\label{decomp}\\
 C^{II}(z)&=&\ll\hat{K}_x^{(0)};\delta\hat{K}_y\gg^{0(0)}_z\,\,\,,
\label{2cf}\\
 C^{III}(z)&=&\ll\hat{K}_x^{(0)};\hat{K}_y^{(0)}\gg^{0(1)}_z\,\,\,.
\label{3cf}
\end{eqnarray}
Obviously, it is sufficient to calculate the correlation function 
consisting of operators (\ref{basicblock}) within the tight binding 
model (\ref{hopping}), however, to first order in the magnetic field.
But first, we note that the functions (\ref{2cf}) and (\ref{3cf}) are
two- and three-point correlation functions, respectively. This is
explicitly seen within the Matsubara representation where the 
expansion of $\ll\hat{K}_x;\hat{K}_y\gg_z^0$ up to first order in the 
``perturbation'' (\ref{perturbation}) yields:
\begin{eqnarray}
 C^{II}(i\omega_m)&=&-\frac{1}{\beta}\int_0^{\beta}
       \int_0^{\beta}d\tau\,d\tau'\,<T_{\tau}\left\{\hat{K}_x^{(0)}(\tau)
                      \,\delta\hat{K}_y(\tau')\right\}>_0^{(0)}\;
                      e^{i\omega_m(\tau-\tau')}\;,\\
 C^{III}(i\omega_m)&=&\frac{1}{\beta}\int_0^{\beta}\int_0^{\beta}
         \int_0^{\beta}
            d\tau\,d\tau'\,d\tau''\,<T_{\tau}\left\{\hat{K}_x^{(0)}(\tau)
         \,\hat{K}_y^{(0)}(\tau')\delta\hat{T}(\tau'')\right\}>_0^{(0)}\;
           e^{i\omega_m(\tau-\tau')}\,.
\end{eqnarray}

\subsection{Expansion to first order in the magnetic field}

As already mentioned, our next goal is to calculate the correlation 
function generated by the operators (\ref{basicblock}) up to first 
order in the magnetic field. This is accomplished by means of its
equation of motion with respect to the tight binding hamiltonian
(\ref{hopping}):
\begin{eqnarray}
  (z+\epsilon_{\vec{k}-\vec{q}}-\epsilon_{\vec{k}}
    &+&\epsilon_{\vec{k}_1}-\epsilon_{\vec{k}_2})
  \ll\hat{A}^{\sigma}_{\vec{k}_1,\vec{k}_2|\vec{k}|\vec{q}}\;;\;
    \hat{A}^{\sigma'}_{\vec{k}'_1,\vec{k}'_2|\vec{k}'|\vec{q'}}
  \gg_z\nonumber\\
 &=&<[\hat{A}^{\sigma}_{\vec{k}_1,\vec{k}_2|\vec{k}|\vec{q}}\;,\;
      \hat{A}^{\sigma'}_{\vec{k}'_1,\vec{k}'_2|\vec{k}'|\vec{q'}}
    ]>-
  \ll[\,\delta\hat{T}\,,\;
    \hat{A}^{\sigma}_{\vec{k}_1,\vec{k}_2|\vec{k}|\vec{q}}\;]\;;\;
    \hat{A}^{\sigma'}_{\vec{k}'_1,\vec{k}'_2|\vec{k}'|\vec{q'}}
  \gg_z^{(0)}\,\,\,.
\label{eom}
\end{eqnarray}
Here and in the following, the index $0$ is omitted. The expansion
of the expectation value on the Rhs.\ with respect to the
magnetic field is standard \cite{fetter} and yields:
\begin{eqnarray}
 <[\hat{A}^{\sigma}_{\vec{k}_1,\vec{k}_2|\vec{k}|\vec{q}}&,&
   \hat{A}^{\sigma'}_{\vec{k}'_1,\vec{k}'_2|\vec{k}'|\vec{q'}}
 ]>\nonumber\\
   &=&<[\hat{A}^{\sigma}_{\vec{k}_1,\vec{k}_2|\vec{k}|\vec{q}}\;,\;
        \hat{A}^{\sigma'}_{\vec{k}'_1,\vec{k}'_2|\vec{k}'|\vec{q'}}
      ]>^{(0)}
   -<\hat{v}\;
      [\hat{A}^{\sigma}_{\vec{k}_1,\vec{k}_2|\vec{k}|\vec{q}}\;,\;
       \hat{A}^{\sigma'}_{\vec{k}'_1,\vec{k}'_2|\vec{k}'|\vec{q'}}
    ]>^{(0)}\label{expansion}\,\,,
\label{expansion}
\end{eqnarray}
where $\hat{v}$ arises from the expansion of the $S$-matrix related to
the ``perturbation'' (\ref{perturbation}) and is therefore given by
\begin{equation}
 \hat{v}=\int_0^{\beta}d\tau\;e^{\tau(\hat{T}^{(0)}-\mu\hat{N})}\;
             \delta\hat{T}\;e^{-\tau(\hat{T}^{(0)}-\mu\hat{N})}\,\,.
\end{equation}
With the representation of $\delta\hat{T}$ in terms of Blochoperators,
\begin{equation}
 \delta\hat{T}=-ietH\sum_{\vec{k}_1\vec{k}_2\sigma}D(\vec{k}_1,\vec{k}_2)
         c_{\vec{k}_1\sigma}^+c_{\vec{k}_2\sigma\;,}
\label{perturbedT}
\end{equation}
where the matrix $D(\vec{k}_1,\vec{k}_2)$ is also defined and further 
evaluated in the appendix (cf.\ Eq.\ (\ref{DDD})), we find more
explicitly: 
\begin{equation}
 \hat{v}=-ietH\sum_{\vec{k}_1\vec{k}_2\sigma}D(\vec{k}_1,\vec{k}_2)
     \frac{e^{\beta(\epsilon_{\vec{k}_1}-\epsilon_{\vec{k}_2})}-1}
                       {\epsilon_{\vec{k}_1}-\epsilon_{\vec{k}_2}}
         c_{\vec{k}_1\sigma}^+c_{\vec{k}_2\sigma\;.}
\label{fromSmatrix}
\end{equation}
Inserting the expansion (\ref{expansion}) into Eq.\ (\ref{eom}), 
we obtain the following zeroth and first order terms for the 
correlation function to be determined:
\begin{eqnarray}
 \ll\hat{A}^{\sigma}_{\vec{k}_1,\vec{k}_2|\vec{k}|\vec{q}}\;;\;
    \hat{A}^{\sigma'}_{\vec{k}'_1,\vec{k}'_2|\vec{k}'|\vec{q'}}
  \gg_z^{(0)}
 &=&\frac{<[
     \hat{A}^{\sigma}_{\vec{k}_1,\vec{k}_2|\vec{k}|\vec{q}}\;,\;
     \hat{A}^{\sigma'}_{\vec{k}'_1,\vec{k}'_2|\vec{k}'|\vec{q'}}
     ]>^{(0)}}{z+\epsilon_{\vec{k}-\vec{q}}-\epsilon_{\vec{k}}
                  +\epsilon_{\vec{k}_1}-\epsilon_{\vec{k}_2}}\;\;
\label{bb2cf},\\
 \ll\hat{A}^{\sigma}_{\vec{k}_1,\vec{k}_2|\vec{k}|\vec{q}}\;;\;
    \hat{A}^{\sigma'}_{\vec{k}'_1,\vec{k}'_2|\vec{k}'|\vec{q'}}
  \gg_z^{(1)}
 &=&-\frac{<\hat{v}\;
      [\hat{A}^{\sigma}_{\vec{k}_1,\vec{k}_2|\vec{k}|\vec{q}}\;,\;
       \hat{A}^{\sigma'}_{\vec{k}'_1,\vec{k}'_2|\vec{k}'|\vec{q'}}
      ]>^{(0)}}{z+\epsilon_{\vec{k}-\vec{q}}-\epsilon_{\vec{k}}
                +\epsilon_{\vec{k}_1}-\epsilon_{\vec{k}_2}}\nonumber\\
 &&-\frac{
  \ll[\,\delta\hat{T}\,,\;
    \hat{A}^{\sigma}_{\vec{k}_1,\vec{k}_2|\vec{k}|\vec{q}}\;]\;;\;
    \hat{A}^{\sigma'}_{\vec{k}'_1,\vec{k}'_2|\vec{k}'|\vec{q'}}
  \gg_z^{(0)}
         }{z+\epsilon_{\vec{k}-\vec{q}}-\epsilon_{\vec{k}}
            +\epsilon_{\vec{k}_1}-\epsilon_{\vec{k}_2}}
\label{1bb3cf}\;\;.
\end{eqnarray}
The second term on the Rhs.\ of Eq.\ (\ref{1bb3cf}) still contains a
correlation function. Fortunately, this function is related to 
the hamiltonian (\ref{freehopping}) without magnetic field thus being
directly reducible to expectation values by means of its equation of
motion:
\begin{equation}
 \ll[\,\delta\hat{T}\,,\;
    \hat{A}^{\sigma}_{\vec{k}_1,\vec{k}_2|\vec{k}|\vec{q}}\;]\;;\;
    \hat{A}^{\sigma'}_{\vec{k}'_1,\vec{k}'_2|\vec{k}'|\vec{q'}}
  \gg_z^{(0)}
 =\frac{
      <[\,[\,\delta\hat{T}\,,\;
        \hat{A}^{\sigma}_{\vec{k}_1,\vec{k}_2|\vec{k}|\vec{q}}\;]\;,\;
        \hat{A}^{\sigma'}_{\vec{k}'_1,\vec{k}'_2|\vec{k}'|\vec{q'}}
      ]>_z^{(0)}
    }{z-\epsilon_{\vec{k}'-\vec{q}'}+\epsilon_{\vec{k}'}
       -\epsilon_{\vec{k}'_1}+\epsilon_{\vec{k}'_2}}\;\;.
\label{2bb3cf}
\end{equation}
In summary, the problem of calculating the memory function
(\ref{memory1}) to leading order in the magnetic field has been 
reduced to the calculation of 
expectation values within a free tight binding model without
magnetic field: The relevant information is contained in the equations 
(\ref{freeK}), (\ref{perturbedK})-(\ref{3cf}) and (\ref{bb2cf})-
(\ref{2bb3cf}). The rather cumbersome calculations are roughly sketched
out in the appendix. We write the memory function contribution to the
Hall constant (cf.\ Eq.\ (\ref{hall})) as follows:
\begin{equation}
 \delta R_H(z)=\frac{1}{e}\left(\frac{U}{2ta_n^0}\right)^2
               \left(m^{II}(z)+m^{III}(z)\right)\;,
\label{RHz}
\end{equation}
where $a_n$ is the amplitude of a nearest neighbour hop and is 
related to the static susceptibility (\ref{staticsus}) via
$\chi^0=4tNa_n$. This in turn implies:
\begin{equation}
 a_n=\frac{1}{2N}\sum_{\vec{k}\sigma}\cos k_x\,n_{\vec{k}\sigma}\,.
\label{a_n}
\end{equation}
$m^{II}(z)$ and $m^{III}(z)$ arise from the two- and three-point 
correlation functions, respectively and are represented with regard to 
the further strategy as energy integrals:
\begin{eqnarray}
 m^{II}(z)&=&-\frac{t^2}{d}\int d\epsilon_1 \int d\epsilon_2
               \int d\epsilon'_1 \int d\epsilon'_2\;
            I(z|\epsilon_1,\epsilon_2|\epsilon'_1,\epsilon'_2)
            L^{II}(\epsilon_1,\epsilon_2|\epsilon'_1,\epsilon'_2)\;,
\label{mII}
\\
 m^{III}(z)&=&\frac{t^3}{\sqrt{d}}(Q(z)+Q(-z))\,,
\label{mIII}
\\
 Q(z)&=&\int d\epsilon \int d\epsilon_1 \int d\epsilon_2
               \int d\epsilon'_1 \int d\epsilon'_2\;
  \frac{I(z|\epsilon,\epsilon_2|\epsilon'_1,\epsilon'_2)
       -I(z|\epsilon_1,\epsilon_2|\epsilon'_1,\epsilon'_2)}
       {\epsilon-\epsilon_1
  }\nonumber\\
 &\times&
   L^{III}(\epsilon|\epsilon_1,\epsilon_2|\epsilon'_1,\epsilon'_2)\;.
\label{QQQ}
\end{eqnarray}
The integrands feature the following abbreviations:
\begin{eqnarray}
 I(z|\epsilon_1,\epsilon_2|\epsilon'_1,\epsilon'_2)&\equiv&
  \frac{f(\epsilon_1)f(\epsilon_2)
          \left[1-f(\epsilon'_1)\right]\left[1-f(\epsilon'_2)\right]
       -\left[1-f(\epsilon_1)\right]\left[1-f(\epsilon_2)\right]
          f(\epsilon'_1)f(\epsilon'_2)
  }{(\epsilon_1+\epsilon_2-\epsilon'_1-\epsilon'_2)
  (z+\epsilon_1+\epsilon_2-\epsilon'_1-\epsilon'_2)}\;,
\label{Izes}
\\
 \frac{L^{II}(\epsilon_1,\epsilon_2|\epsilon'_1,\epsilon'_2)}{d}
 &\equiv&\left.\right<
   \delta(\epsilon_1-\epsilon_{\vec{k}_1})
   \delta(\epsilon_2-\epsilon_{\vec{k}_2})
   \delta(\epsilon'_1-\epsilon_{\vec{k}'_1})
   \delta(\epsilon'_2-\epsilon_{\vec{k}'_2})\nonumber\\
 &\times&\{\cos k_{1x}+\cos k'_{1x}-[\sin k_{1x}-\sin k'_{1x}]
  \,{\cal P}\cot(\frac{k_{1x}+k_{2x}-k'_{1x}-k'_{2x}}{2})\}\nonumber\\
 &\times&\{\cos k_{1y}+\cos k_{2y}-\cos k'_{1y}-\cos k'_{2y}\}\nonumber\\
 &\times&2\pi\,\delta(k_{1y}+k_{2y}-k'_{1y}-k'_{2y})
 \left>_{\vec{k}_1\vec{k}_2\vec{k}'_1\vec{k}'_2\;,}\right.
\label{LII}
\\
 \frac{L^{III}(\epsilon|\epsilon_1,\epsilon_2|\epsilon'_1,\epsilon'_2)}
      {-2\sqrt{d}
 }&\equiv&\left.\right<
   \delta(\epsilon-\epsilon_{\vec{k}'_1+\vec{k}'_2-\vec{k}_2})
   \delta(\epsilon_1-\epsilon_{\vec{k}_1})
   \delta(\epsilon_2-\epsilon_{\vec{k}_2})
   \delta(\epsilon'_1-\epsilon_{\vec{k}'_1})
   \delta(\epsilon'_2-\epsilon_{\vec{k}'_2})\nonumber\\
 &\times&\{\cos k_{1x}+\cos k_{2x}+\cos k'_{1x}+\cos
 k'_{2x}\nonumber\\
 &-&[\sin k_{1x}+\sin k_{2x}-\sin k'_{1x}-\sin k'_{2x}]
  \,{\cal P}\cot(\frac{k_{1x}+k_{2x}-k'_{1x}-k'_{2x}}{2})\}\nonumber\\
 &\times&\{\sin k_{1y}+\sin k_{2y}-\sin k'_{1y}-\sin k'_{2y}\}\nonumber\\
 &\times&\sin k_{1y}2\pi\,\delta(k_{1y}+k_{2y}-k'_{1y}-k'_{2y})
 \left>_{\vec{k}_1\vec{k}_2\vec{k}'_1\vec{k}'_2\;,}\right.
\label{LIII}
\end{eqnarray}
and $f(\epsilon)\equiv1/(\exp(\epsilon-\mu)+1)$ is the Fermi function.
$<\ldots>_{\vec{k}}$ denotes the average over the first Brillouin
zone, i.e.\ $\int d^d k/(2\pi)^d(\ldots)$.
Note, that the last two equations (\ref{LII}) and (\ref{LIII}) reflect
the gauge fixation (\ref{gauge}): In the $y$-direction, crystal
momentum is conserved which is ensured by the $\delta$-functions while 
the $x$-components of the Bloch vectors are coupled more complicatedly.
In principle, we could do the momentum integrations numerically for
a two-dimensional lattice and for given sets of external parameters 
temperature $T$, doping $\delta$ and frequency $\omega$. 
However, we may carry on our analysis a little bit by invoking a limit 
pioneered by Metzner and Vollhardt in the context of strongly 
correlated electrons \cite{metzner}, namely the limit of infinite 
spatial dimensions. In this limit, the momentum integrals decouple and
we are left with energy integrals over smooth functions. This
procedure will be discussed in the next subsection.

\subsection{The limit of infinite lattice dimensions}

We may question the relevance of this limit, since the important
physics of the high-$T_c$ superconductors is known to take place in 
Cu-O planes. Many authors have addressed this issue and much evidence 
has been revealed in favour of the relevance of this limiting
procedure even for two-dimensional systems, see e.g.\ Ref.\
\cite{pruschke1}. Instead of immersing ourselves in this debate, we
take the following point of view: The main reasons behind the
anomalous properties of the high-$T_c$ materials seem to be, firstly, the
strong electronic correlations and, secondly, the two-dimensionality 
of the relevant Cu-O planes. Taking the limit $d\rightarrow\infty$ 
helps us to separate the impact of the correlations and to suppress 
effects of low-dimensionality like e.g.\ van Hove 
singularities. In this sense, the limit $d\rightarrow\infty$ is 
interesting in itself.

For our problem, the most important aspect of the limit
$d\rightarrow\infty$ is the following: For the Hubbard model 
to retain its nontrivial dynamics, the parameter $t$ has to be scaled 
properly with $d$ according to
\begin{equation}
 2t=\frac{t^*}{\sqrt{d}}\;
\label{scaling}
\end{equation}
(in this subsection, we set $t^*\equiv1$). Only then does the Hubbard model
capture simultaneously the itinerant and local aspect introduced by 
the hopping and interaction term, respectively. On the other hand,
we are tempted to conclude from the scaling (\ref{scaling}) that any 
transport stops to be possible in $d=\infty$. In fact, a more thorough
investigation shows that the longitudinal and the Hall conductivity are
of order $1/d$ and $1/d^2$, respectively. But this, in turn, implies 
that the Hall constant remains finite in $d\rightarrow\infty$. In the
following, all we need to know is how to calculate averages over
Brillouin zones of the type (\ref{LII}) and (\ref{LIII}). The
corresponding procedure is explained in the appendix and enables us
also to calculate simpler quantities as the density of states
$D(\epsilon)$ of the band model (\ref{freehopping}), the nearest 
neigbor hopping amplitude (\ref{a_n}) and the amplitude of a hop 
diagonally across the unit cell, i.e.\
\begin{equation}
 a_d=\frac{1}{2N}\sum_{\vec{k}\sigma}\cos k_x\cos k_y\,n_{\vec{k}\sigma}\,.
\label{a_d}
\end{equation}
In the case of the amplitudes (\ref{a_n}) and (\ref{a_d}), the
functions
\begin{eqnarray}
 A(\epsilon)&\equiv&<\cos k_x\,\delta(\epsilon-\epsilon_{\vec{k}})
                    >_{\vec{k}}=
     -\frac{\epsilon}{\sqrt{d}}D(\epsilon)\;,
\label{A_e}\\
 B(\epsilon)&\equiv&<\cos k_x\cos
                      k_y\,\delta(\epsilon-\epsilon_{\vec{k}})
                    >_{\vec{k}}=
     \frac{1}{d}(\epsilon^2-\frac{1}{2})D(\epsilon)
\label{B_e}
\end{eqnarray}
come into play. Essentially, they are Gaussians multiplied with the
first and second Hermitian polynomial, respectively, since they are
derivatives of the Gaussian density of states:
\begin{equation}
 D(\epsilon)=\frac{1}{\sqrt{\pi}}\,e^{-\epsilon^2}\;.
\label{DOS}
\end{equation}
The functions (\ref{LII}) and (\ref{LIII}) are found to be combinations
of the functions (\ref{A_e}) and (\ref{B_e}) and may be written as
\begin{eqnarray}
 L^{II}(\epsilon_1,\epsilon_2|\epsilon'_1,\epsilon'_2)
 &=&D(\epsilon_1)D(\epsilon_2)D(\epsilon'_1)D(\epsilon'_2)
     \{\epsilon_1^2+\epsilon_1\epsilon_2+\epsilon_2\epsilon'_1
     -{\epsilon'_1}^2-\epsilon'_1\epsilon'_2-\epsilon'_2\epsilon_1\}\;,
\label{LIIeva}
\\
 L^{III}(\epsilon|\epsilon_1,\epsilon_2|\epsilon'_1,\epsilon'_2)
 &=&D(\epsilon)D(\epsilon_1)D(\epsilon_2)D(\epsilon'_1)D(\epsilon'_2)
   \{-\epsilon+\epsilon_1+\epsilon_2+\epsilon'_1+\epsilon'_2\}\;.
\label{LIIIeva}
\end{eqnarray}
In order to handle the singularity of the Hall constant in the empty 
band limit correctly, we shall discuss the perturbation-theoretical 
results for the Hall constant normalized to its $U\rightarrow0$ limit:
\begin{equation}
 \frac{R_H(z,U)}{R_{H0}}=1+U^2(K^{\infty}+K^{II}(z)+K^{III}(z))\;.
\label{result}
\end{equation}
Here, the $U=0$ Hall constant is given by
\begin{equation}
 R_{H0}=\frac{1}{e}\frac{2a_{d0}}{(2a_{n0})^2}
\label{freeHall}
\end{equation}
(the subscripts $0$ indicate $U=0$ as above) and $K^{II}(z)$ 
and $K^{III}(z)$ arise from the functions (\ref{mII}) and
(\ref{mIII}), e.g.\ $K^{II}(z)=(2d/a_{d0})m^{II}(z)$. $K^{\infty}$ is
the perturbation-theoretical contribution of the infinite frequency
Hall constant (\ref{R_Hinfty}). Since the latter is given by 
$R_H^{\infty}=(1/e)(2a_d/(2a_n)^2)$,
it suffices to calculate the density $n_{\vec{k}\sigma}$ to second
order in $U$. The derivation is standard and will therefore not be
given here. Due to symmetries of the expression (\ref{LIIIeva}), the 
function (\ref{QQQ}) may be simplified by means of various
redefinitions of the integration variables:
\begin{eqnarray}
 Q(z)&=&\int d\epsilon_1 \int d\epsilon_2
        \int d\epsilon'_1 \int d\epsilon'_2\;
        D(\epsilon_1)D(\epsilon_2)D(\epsilon'_1)D(\epsilon'_2)\nonumber\\
 &\times&I(z|\epsilon_1,\epsilon_2|\epsilon'_1,\epsilon'_2)\;
        \{\epsilon_2+\epsilon'_1+\epsilon'_2\} 
      \int d\epsilon 
        \frac{D(\epsilon_1+\epsilon)-D(\epsilon_1-\epsilon)}{\epsilon}\;.
\label{Qeva}
\end{eqnarray}
The terms of Eq.\ (\ref{result}) may now be evaluated numerically in 
the limit $\omega\rightarrow0$ on the basis of Eqs.\ (\ref{mII}), 
(\ref{mIII}) and (\ref{Qeva}) along with the definition (\ref{Izes}) 
and the results (\ref{LIIeva}) and (\ref{LIIIeva}). But first, we
check whether the Hall constant (\ref{result}) reduces to the
familiar expression $1/ne$, provided the electron density
$n=(1/N)\sum_{\vec{k}\sigma}n_{\vec{k}\sigma}$ is very low. We
concentrate on zero temperature, where we find the Fermi energy to be
given by $n=1+\mbox{erf}(\epsilon_F)$. This implies
$\epsilon_F\rightarrow-\infty$ in the empty band limit. Then, the
corrections on the Rhs.\ of Eq.\ (\ref{result}) vanish and the Hall
constant is given by Eq.\ (\ref{freeHall}). With
$2a_{n0}=D(\epsilon_F)/\sqrt{d}$ and
$2a_{d0}=-\epsilon_FD(\epsilon_F)/d$, we find in fact for
$n\rightarrow0$: 
\begin{equation}
 R_H=-\frac{\epsilon_F}{e\,D(\epsilon_F)}\simeq\frac{1}{en}\,.
\label{familiarHall}
\end{equation}

\subsection{Numerical results}

\begin{figure}
 \begin{center}
  \mbox{\psfig{figure=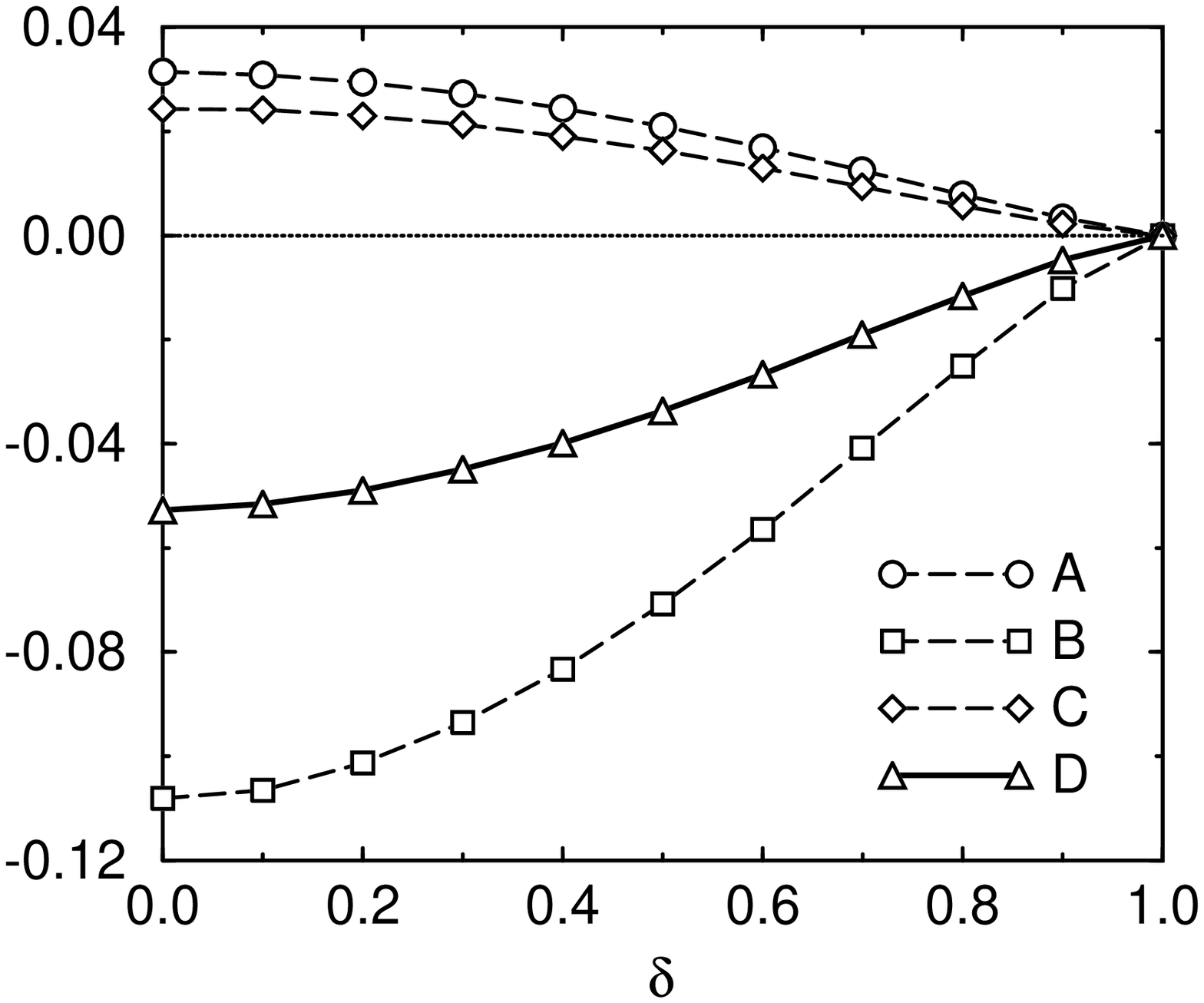,width=11cm,height=8cm}}
 \end{center}
 \caption{\sf The corrections of Eq.\ (\ref{result}) at $T=0$. The
              curves A, B, C, and D represent the functions 
              $K^{\infty}(\delta)$, $K^{II}(\delta)$,
              $K^{III}(\delta)$, and their sum, respectively.}
          \label{corrections}
\end{figure}

First of all, we discuss the relative importance of the terms
appearing on the Rhs.\ of Eq.\ (\ref{result}). Fig.\
\ref{corrections} shows their doping dependence at $T=0$ (dashed
lines) and that of their sum (solid line), where the doping parameter
is defined as $\delta=1-n$. All functions vanish in the empty band 
limit ($\delta\rightarrow1$) and exhibit monotonic behaviour with
decreasing doping. This reflects the fact, that the suppression of 
doubly occupied sites introduced by the Hubbard repulsion 
(\ref{interaction}) becomes more effective with increasing electron 
density. As for the signs of the three contributions, only that of 
the two-point correlation function is negative. This, however, is 
sufficient to render the sum of all terms negative (solid line in 
Fig.\ \ref{corrections}). This remains valid at finite temperatures:  
The term $K^{\infty}$ is positive for all temperatures and dopings 
considered, but is always overcompensated by the memory function 
contribution $K^{II}(i0^+)+K^{III}(i0^+)$. Thus, our 
perturbation-theoretical results clearly indicate the tendency of the 
Hall constant to change its sign {\em below} half filling. However,
for this to happen, the memory function contribution must be taken
into account. 

To study the doping and temperature dependence of this precursor
effect in greater detail, we shall extrapolate Eq.\ (\ref{result}) to
correlation strengths $U$ big enough for the Hall constant to exhibit
a sign change. Ultimately, we fix $U$ such that this sign change
occurs in the parameter regime observed experimentally in the case of
the compound La$_{2-\delta}$Sr$_{\delta}$CuO$_4$. We shall measure $U$
in terms of the bare band width $W$, which may be chosen, due to Eq.\ 
(\ref{DOS}), as $W=2t^*$ (from now on, the hopping parameter $t^*$ is 
explicitly taken into account). Then, it follows from Eq.\ 
(\ref{result}) and Fig.\ \ref{corrections}, that the $T=0$ threshold 
$U$, above which the Hall constant becomes positive, is $U=2.18 W$
at half filling ($\delta=0$) and increases monotonically with
increasing doping and ultimately diverges in the empty band limit 
$\delta\rightarrow1$. 

Before proceeding, we touch upon the issue of how to relate our 
theoretical results to experimental measurements. Firstly, the
$d=\infty$ hopping parameter $t^*$ has been estimated crudely as
$0.2t^*\sim500K$ \cite{pruschke2}. Secondly, we shall express the Hall
constant in units that allow direct comparison with experimental
results. This requires that charge carrier densities are taken with
respect to the volume of a unit cell. On the other hand, the electron 
density $n$, appearing in Eq.\ (\ref{familiarHall}), denotes the
average number of electrons per lattice site. From a theoretical point
of view, this definition is convenient since it is independent of the 
lattice dimension $d$ and remains meaningful in the limit 
$d\rightarrow\infty$. Therefore, in order to compare our theoretical 
results with measurements on a certain cuprate, we have to multiply
the Hall constant of Eq.\ (\ref{result}) by $\Omega/\nu$, where 
$\Omega$ is the volume of a unit cell and $\nu$ the number of Cu ions 
therein. In the case of La$_{2-\delta}$Sr$_{\delta}$CuO$_4$, 
$\Omega=186\,\mbox{{\AA}}^3$ and $\nu=2$.

\begin{figure}
 \begin{center}
  \mbox{\psfig{figure=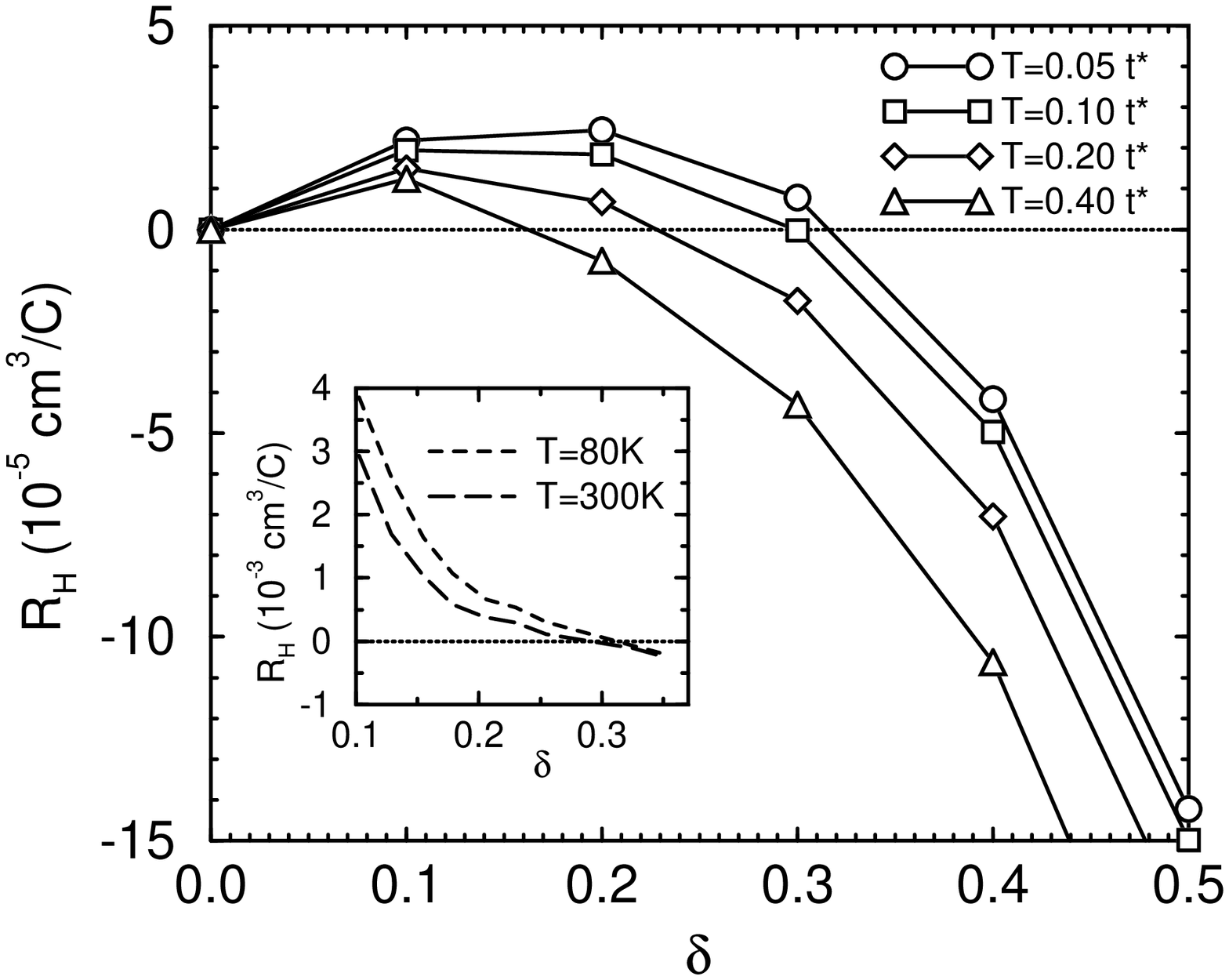,width=11cm,height=8cm}}
 \end{center}
 \caption{\sf Hall constant as a function of doping for 
              $U=2.3\,W$. Inset: Data for polycrystalline 
              samples of La$_{2-\delta}$Sr$_{\delta}$CuO$_4$  
              taken from Ref.\ {\protect\cite{takagi}}.}
 \label{dopcomp}
\end{figure}

\begin{figure}
 \begin{center}
  \mbox{\psfig{figure=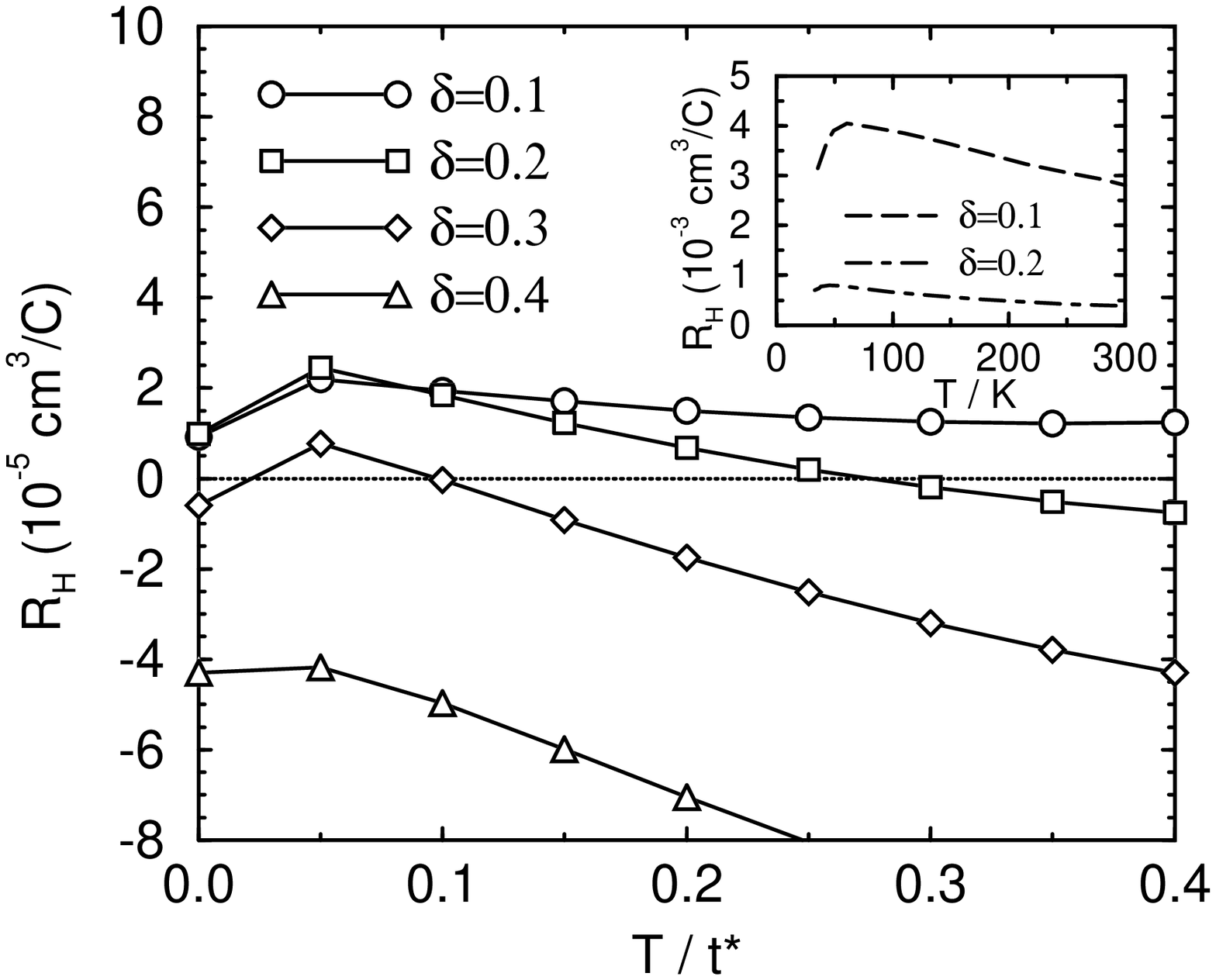,width=11cm,height=8cm}}
 \end{center}
 \caption{\sf Hall constant as a function of temperature 
              for $U=2.3\,W$. Inset: Data for polycrystalline 
              samples of La$_{2-\delta}$Sr$_{\delta}$CuO$_4$  
              taken from Ref.\ {\protect\cite{hwang}}.}
 \label{temcomp}
\end{figure}

After these preliminary remarks, we proceed with the investigation
of the intermediate correlation regime, where a sign change is
possible. Fig.\ \ref{dopcomp} shows the doping dependent Hall constant
for several temperatures and the choice $U=2.3W$ (solid lines) as well
as two experimental curves for polycrystalline samples of
La$_{2-\delta}$Sr$_{\delta}$CuO$_4$ taken from Ref.\ \cite{takagi} 
(inset). We see that the sign change occurs close to
$\delta\approx0.3$ for temperatures below $300K$, in agreement with
experiment. For lower dopings, our Hall constant exhibits a
maximum and ultimately vanishes at half filling, irrespective of 
temperature. This reflects the fact that, in our 
perturbation-theoretical result (\ref{result}), the Hall constant of
the bare band is merely renormalized by a finite factor. Such a factor
may change an overall sign but never can turn a vanishing quantity
into a nonzero one. Thus, our results fail to account for the observed
$1/\delta$ law of the Hall constant near half filling. And that is why
our perturbation-theoretical curves cross the zero line (dotted line
in Fig.\ \ref{dopcomp}) with slopes that are two orders of magnitude 
smaller than those of the experimental curves. Apart from this 
deficiency of perturbation theory, the other dependences are in 
qualitative agreement with experiment. In Fig.\ \ref{temcomp}, the 
temperature dependence of the Hall constant is shown for various 
dopings within the range $0.1\le\delta\le0.4$ (solid lines) and 
compared to experimental results from Ref.\ \cite{hwang}, again for
polycrystalline samples of La$_{2-\delta}$Sr$_{\delta}$CuO$_4$ (dashed
lines in the inset). Despite the already mentioned difference in the
order of magnitude of the Hall constant, our curves display the same 
features as the experimental ones: A maximum at low temperatures 
followed by a regime in which the Hall constant decreases
monotonically up to unusually high temperatures. We have not been able
to determine the exact location of the maximum due to numerical 
difficulties at nonzero temperatures below $0.05 t^*$. Experimentally,
it occurs above $T_c$. The fact that it appears within the Hubbard
model suggests that it is not related to the onset of superconducting 
correlations. This is further supported by comparing the data of
``90-K'' and ``60-K'' YBa$_2$Cu$_3$O$_{6+x}$ \cite{harris}: In this 
compound, the location of the maximum in $R_H(T)$ does not depend on
the doping values corresponding to the range $60K<T_c<90K$. As for the
decrease of the Hall constant as a function of temperature, it is 
experimentally found to be most pronounced at optimal doping, i.e.\ 
$\delta=0.15$ in the case of La$_{2-\delta}$Sr$_{\delta}$CuO$_4$. Our 
results exaggerate the doping range where this decrease is markedly 
visible. At least, Fig.\ \ref{temcomp} shows that the decrease is least
pronounced for the curve corresponding to the lowest doping value,
$\delta=0.1$. Furthermore, at high dopings, where the memory function 
contribution becomes unimportant, the Hall constant becomes almost 
temperature independent.

What about the observed quadratic dependence of the Hall angle on 
temperature for small dopings \cite{harris,chien,hwang}? This law
cannot be verified on the basis of Eq.\ (\ref{result}) alone,
although, it cannot be falsified either. To make a check on this law, 
the longitudinal conductivity is needed as well. In principle, the
calculation of this quantity can be done along the same lines leading 
to Eq.\ (\ref{result}) and is left for future work.

\section{Conclusions}
In summary, we have devised a memory function approach to the Hall
constant in strongly correlated electron systems, which enables us to
cope directly with the Hall resistivity. As a first step, its
usefulness was demonstrated in a perturbation-theoretical treatment
within the single-band Hubbard model. To obtain a regular expression 
for the memory function contribution for all frequencies, we assumed 
that the subspace of the operator space spanned by the current 
operators $\hat{J}_x$ and $\hat{J}_y$ is invariant with respect to 
actions of the unperturbed Liouville operator. This approximation was 
shown to become exact in the continuum limit. Furthermore, it affects 
only the properties of the unperturbed system, i.e.\ the tight-binding
electrons. Therefore, we do not expect the omitted terms to change the
physics in an essential way. On the basis of this approximation, the 
memory function was calculated to leading order in the correlation 
strength $U$ and shown to decompose into a two- and three-point 
correlation function, when expanded to first order in the magnetic
field. The complicated expressions obtained for these two functions
were simplified considerably by invoking the limit of infinite spatial
dimensions. While this approximation still catches the impact of the 
correlations, it smoothes out effects of low dimensionality. Except
for the doping dependence of the Hall constant in the vicinity of half
filling, we have been able to reproduce the unusual experimental
findings in connection with high-$T_c$ superconductors as 
La$_{2-\delta}$Sr$_{\delta}$CuO$_4$. In particular, we could explain 
the sign change of the Hall constant as a function of doping and its 
decrease as a function of temperature up to unusually high
temperatures. However, we had to chose $U=2.3W$ ($W$: bare band
width), which is, strictly speaking, outside the perturbative regime. 
Since the cuprates are believed to undergo a transition from a Fermi 
liquid to a strong-correlation regime when the doping approaches its 
optimal value from the overdoped side \cite{ong1}, it is not
astonishing that the $1/\delta$ law in the vicinity of the mother 
compound cannot be described within perturbation theory. A treatment
of the Hall effect in the opposite limit of strong correlations is in 
progress \cite{lange}. 

\acknowledgments
The author is grateful to P.\ W\"olfle for many stimulating
discussions. This work has been supported by the
Landesforschungsschwerpunktprogramm and the Sonderforschungsbereich
195. 

\begin{appendix}

\section{Lattice sums over nearest neighbours}

Since our gauge choice, cf.\ Eq.\ (\ref{gauge}), permits periodic
boundary conditions in the $y$-direction only, we must carry out all 
sums over nearest neighbours according to the following formula:
\begin{equation}
 \sum_{\vec{R},\vec{R}+\vec{\delta}}
      H(\vec{R},\vec{R}+\vec{\delta})
   =\sum_{R_x}\sum_{R_y}\sum_{\delta_y}
         H(\vec{R},\vec{R}+\delta_y\,\hat{y})
   +\sum_{R_x+\delta_x\atop R_x}\sum_{R_y}
         H(\vec{R},\vec{R}+\delta_x\,\hat{x})\;.
\label{summation}
\end{equation}
In the first term on the Rhs., we may carry out the sum over $R_y$ and
$\delta_y$ independently. As for the second term on the Rhs., we must
make sure that $R_x$ and $R_x+\delta_x$ are neighbouring elements of 
the set $\{R_x^{min},R_x^{min}+1,\ldots,R_x^{max}\}$ with $R_x^{max}
\equiv R_x^{min}+N_x-1$ as explained in the text following Eq.\ 
(\ref{perturbedcurrent}).

First of all, the matrices appearing in Eqs.\ (\ref{freeK}), 
(\ref{perturbedK}) and (\ref{perturbedT}) are defined as follows: 
\begin{eqnarray}
 B_{\nu}(\vec{k}_1,\vec{k}_2)&=&\frac{it}{N}
     \sum_{\vec{R},\vec{R}+\vec{\delta}}\delta_{\nu}\,
     e^{-i\vec{k}_1(\vec{R}+\vec{\delta})+i\vec{k}_2\vec{R}}\;\;,
\label{Bsubnu}
\\
 C(\vec{k}_1,\vec{k}_2)&=&\frac{1}{N}
     \sum_{\vec{R},\vec{R}+\vec{\delta}}(\delta_y)^2R_x\,
     e^{-i\vec{k}_1(\vec{R}+\vec{\delta})+i\vec{k}_2\vec{R}}\;\;,
\label{CCC}
\\
 D(\vec{k}_1,\vec{k}_2)&=&\frac{1}{N}
     \sum_{\vec{R},\vec{R}+\vec{\delta}}\delta_y R_x\,
     e^{-i\vec{k}_1(\vec{R}+\vec{\delta})+i\vec{k}_2\vec{R}}\;\;.
\label{DDD}
\end{eqnarray}
In all these cases, the integrand is proportional to a component
of a nearest neighbour vector, which entails that one of the two terms 
of Eq.\ (\ref{summation}) vanishes. The summation over the
$y$-components $R_y$ and $\delta_y$ is straightforward. In the case of
Eqs.\ (\ref{CCC}) and (\ref{DDD}), the following sum is to be
calculated:
\begin{equation}
 \frac{1}{N_x}\sum_{R_x}
   R_x\,e^{i(k_{2x}-k_{1x})R_x}
   =\delta_{k_{1x}|k_{2x}}R_x^c+
    (1-\delta_{k_{1x}|k_{2x}})\frac{e^{i(k_{2x}-k_{1x})R_x^{min}}}
    {e^{i(k_{2x}-k_{1x})}-1}\;,
\end{equation}
where $R_x^c\equiv(R_x^{min}+R_x^{max})/2$ is the $x$-component of the
lattice's center of gravity. In summary, we obtain the following
results: 
\begin{eqnarray}
 B_x(\vec{k}_1,\vec{k}_2)&=&\delta_{\vec{k}_1|\vec{k}_2}v_x(\vec{k}_1)
     +\delta_{k_{2y}|k_{1y}}\frac{it}{N_x}e^{i(k_{2x}-k_{1x})R_x^{min}}
     \left(e^{ik_{1x}}-e^{-ik_{2x}}\right),
\label{Bsubx}
\\
 B_y(\vec{k}_1,\vec{k}_2)&=&\delta_{\vec{k}_1|\vec{k}_2}v_y(\vec{k}_1),
\label{Bsuby}
\\
 C(\vec{k}_1,\vec{k}_2)&=&2\cos k_{1y}\delta_{k_{1y}|k_{2y}}
    \left(\delta_{k_{1x}|k_{2x}}R_x^c+
    (1-\delta_{k_{1x}|k_{2x}})\frac{e^{i(k_{2x}-k_{1x})R_x^{min}}}
    {e^{i(k_{2x}-k_{1x})}-1}\right),
\label{Cresult}
\\
 D(\vec{k}_1,\vec{k}_2)&=&-2i\sin k_{1y}\delta_{k_{1y}|k_{2y}}
    \left(\delta_{k_{1x}|k_{2x}}R_x^c+
    (1-\delta_{k_{1x}|k_{2x}})\frac{e^{i(k_{2x}-k_{1x})R_x^{min}}}
    {e^{i(k_{2x}-k_{1x})}-1}\right)\,.
\label{Dresult}
\end{eqnarray}
In the following section, we shall see that, once these expressions
are inserted into observable quantities, the final results become
independent of the lattice location.

The commutators (\ref{condition1}) and (\ref{condition2}) are also
calculated using Eq.\ (\ref{summation}).

\section{Evaluation of the two- and three-point correlation function}

We begin by inserting Eqs.\ (\ref{freeK}) and (\ref{perturbedK}) 
into Eqs.\ (\ref{2cf}) and (\ref{3cf}) and by using Eqs.\
(\ref{bb2cf})-(\ref{2bb3cf}) along with Eqs.\
(\ref{Bsubx})-(\ref{Dresult}). Then, the correlation functions 
(\ref{2cf}) and (\ref{3cf}) are given as sums over terms that contain
expectation values with respect to the momentum conserving Hamiltonian
(\ref{freehopping}). Taking into account the corresponding delta
functions, we see that the diagonal elements of the matrices
(\ref{Bsubx}), (\ref{Cresult}) and (\ref{Dresult}) lead to vanishing 
contributions due to symmetry arguments. Moreover, the combination of 
all exponentials whose arguments are proportional to $R^{min}_x$ 
may be replaced by one due to the delta functions, expressing
momentum conservation in the model (\ref{freehopping}). With the help 
of the function
\begin{eqnarray}
 S(\vec{k})&=&N_y\delta_{k_y|0}(1-\delta_{k_x|0})\frac{1}{e^{ik_x}-1}
             \nonumber\\
           &=&-\frac{1}{2}N_y\delta_{k_y|0}\,[1+i{\cal P}\cot(\frac{k_x}{2})]
\end{eqnarray}
and after some straightforward manipulations, we arrive at:
\begin{eqnarray}
 \frac{C^{II}(z)}{-2iHet^2NU^2}&=&\frac{1}{N^4}
  \sum_{\vec{k_1}\vec{k_2}\vec{k}\vec{q}\sigma}\; 
  \sum_{\vec{k'_1}\vec{k'_2}\vec{k'}\vec{q'}\sigma'}\;
  \delta_{k_{2y}|k_{1y}-q_y}\;S(\vec{k'_2}+\vec{q'}-\vec{k'_1})\nonumber\\
 &\times&\left(e^{ik_{1x}}-e^{-i(k_{2x}+q_x)}
                   -e^{i(k_{1x}-q_x)}+e^{-ik_{2x}}\right)
     \left(\cos k'_{1y}-\cos(k'_{1y}-q'_y)\right)\nonumber\\
 &\times&
 \frac{<[
     \hat{A}^{\sigma}_{\vec{k}_1,\vec{k}_2|\vec{k}|\vec{q}}\;,\;
     \hat{A}^{\sigma'}_{\vec{k}'_1,\vec{k}'_2|\vec{k}'|\vec{q'}}
 ]>^{(0)}_0}{z+\epsilon_{\vec{k}-\vec{q}}-\epsilon_{\vec{k}}
                  +\epsilon_{\vec{k}_1}-\epsilon_{\vec{k}_2}}\;\;,
\label{cf2interim}\\
 \frac{C^{III}(z)}{4iHet^3NU^2}&=&P_1(z)+P_2(z)\;,
\label{cf3p1interim}\\
 P_1(z)&=&\frac{1}{N^4}
   \sum_{\vec{k_1}\vec{k_2}\vec{k}\vec{q}\sigma}\; 
   \sum_{\vec{k'_1}\vec{k'_2}\vec{k'}\vec{q'}\sigma'}\; 
   \sum_{\vec{p_1}\vec{p_2}\tau}S(\vec{p_2}-\vec{p_1})\,
   \delta_{\vec{k'_2}+\vec{q'}-\vec{k'_1}|\vec{0}}\;\nonumber\\
 &\times&\sin p_{1y}\;[\sin k'_{1y}-\sin(k'_{1y}-q'_y)]\;
   (1-e^{-iq_x})(e^{ik_{1x}}+e^{-ik_{2x}})\;\nonumber\\
 &\times&
   \frac{e^{\beta(\epsilon_{\vec{p}_1}-\epsilon_{\vec{p}_2})}-1}
        {\epsilon_{\vec{p}_1}-\epsilon_{\vec{p}_2}}\;\;
   \frac{<c_{\vec{p}_1\tau}^+c_{\vec{p}_2\tau}\;
         [\hat{A}^{\sigma}_{\vec{k}_1,\vec{k}_2|\vec{k}|\vec{q}}\;,\;
          \hat{A}^{\sigma'}_{\vec{k}'_1,\vec{k}'_2|\vec{k}'|\vec{q'}}
         ]>^{(0)}_0
        }{z+\epsilon_{\vec{k}-\vec{q}}-\epsilon_{\vec{k}}
           +\epsilon_{\vec{k}_1}-\epsilon_{\vec{k}_2}}\;\;,
\label{cf3p2interim}\\
 P_2(z)&=&\frac{1}{N^4}
   \sum_{\vec{k_1}\vec{k_2}\vec{k}\vec{q}\sigma}\; 
   \sum_{\vec{k'_1}\vec{k'_2}\vec{k'}\vec{q'}\sigma'}\; 
   \sum_{\vec{p_1}\vec{p_2}\tau}S(\vec{p_2}-\vec{p_1})\,\,
   \delta_{\vec{k'_2}+\vec{q'}-\vec{k'_1}|\vec{0}}\;\,\nonumber\\
 &\times&\frac{\sin p_{1y}\;[\sin k'_{1y}-\sin(k'_{1y}-q'_y)]
     (1-e^{-iq_x})(e^{ik_{1x}}+e^{-ik_{2x}})
  }{\left(z+\epsilon_{\vec{k}-\vec{q}}-\epsilon_{\vec{k}}
           +\epsilon_{\vec{k}_1}-\epsilon_{\vec{k}_2}\right)
    \left(z-\epsilon_{\vec{k'}-\vec{q'}}+\epsilon_{\vec{k'}}
                  -\epsilon_{\vec{k}_1'}+\epsilon_{\vec{k}_2'}\right)
   }\nonumber\\
 &\times&
 \left\{\delta_{\tau|\sigma}\left[
   \delta_{\vec{p}_2|\vec{k}_1}
     <[\hat{A}^{\sigma}_{\vec{p}_1,\vec{k}_2|\vec{k}|\vec{q}}\;,\;
       \hat{A}^{\sigma'}_{\vec{k}'_1,\vec{k}'_2|\vec{k}'|\vec{q'}}
     ]>^{(0)}_0\right.\right.\nonumber\\
 &-&\left.\delta_{\vec{p}_1|\vec{k}_2}
     <[\hat{A}^{\sigma}_{\vec{k}_1,\vec{p}_2|\vec{k}|\vec{q}}\;,\;
       \hat{A}^{\sigma'}_{\vec{k}'_1,\vec{k}'_2|\vec{k}'|\vec{q'}}
     ]>^{(0)}_0\right]\nonumber\\
 &+&
   \delta_{\tau|-\sigma}\left[
   \delta_{\vec{p}_2|\vec{k}-\vec{q}}
     <[\hat{A}^{\sigma}_{\vec{k}_1,\vec{k}_2|\vec{k}|\vec{k}-\vec{p}_1}\;,\;
       \hat{A}^{\sigma'}_{\vec{k}'_1,\vec{k}'_2|\vec{k}'|\vec{q'}}
     ]>^{(0)}_0\right.\nonumber\\
 &-&\left.\left.\delta_{\vec{p}_1|\vec{k}}
     <[\hat{A}^{\sigma}_{\vec{k}_1,\vec{k}_2|\vec{p}_2|
                                   \vec{p}_2-\vec{k}+\vec{q}}\;,\;
       \hat{A}^{\sigma'}_{\vec{k}'_1,\vec{k}'_2|\vec{k}'|\vec{q'}}
     ]>^{(0)}_0\right]\right\}\;.
\label{cf3p3interim}
\end{eqnarray}
From the definition (\ref{basicblock}), we find:
\begin{eqnarray}
 [\hat{A}^{\sigma}_{\vec{k}_1,\vec{k}_2|\vec{k}|\vec{q}}\,,\,
  \hat{A}^{\sigma'}_{\vec{k}'_1,\vec{k}'_2|\vec{k}'|\vec{q'}}
 ]&=&\delta_{\sigma'|\sigma}
      c^+_{\vec{k}_1|\sigma}c_{\vec{k}_2|\sigma}
      c^+_{\vec{k}'_1|\sigma}c_{\vec{k}'_2|\sigma}
      \{\delta_{\vec{k}|\vec{k}'-\vec{q}'}
         c^+_{\vec{k}-\vec{q}|-\sigma}c_{\vec{k}'|-\sigma}
       -\delta_{\vec{k}'|\vec{k}-\vec{q}}
         c^+_{\vec{k}'-\vec{q}'|-\sigma}c_{\vec{k}|-\sigma}
      \}\nonumber\\
 &+&\delta_{\sigma'|\sigma}
      \{\delta_{\vec{k}_2|\vec{k}'_1}
         c^+_{\vec{k}_1\sigma}c_{\vec{k}'_2\sigma}
       -\delta_{\vec{k}'_2|\vec{k}_1}
         c^+_{\vec{k}'_1\sigma}c_{\vec{k}_2\sigma}
      \}
      c^+_{\vec{k}'-\vec{q}'|-\sigma}c_{\vec{k}'|-\sigma}
      c^+_{\vec{k}-\vec{q}|-\sigma}c_{\vec{k}|-\sigma}
         \nonumber\\
 &+&\delta_{\sigma'|-\sigma}
      \{\delta_{\vec{k}|\vec{k}'_1}
          c^+_{\vec{k}-\vec{q}|-\sigma}c_{\vec{k}'_2|-\sigma}
          c^+_{\vec{k}_1|\sigma}c_{\vec{k}_2|\sigma}
          c^+_{\vec{k}'-\vec{q}'|\sigma}c_{\vec{k}'|\sigma}
         \nonumber\\
 &&\qquad-\delta_{\vec{k}'|\vec{k}_1}
          c^+_{\vec{k}'-\vec{q}'|\sigma}c_{\vec{k}_2|\sigma}
          c^+_{\vec{k}'_1|-\sigma}c_{\vec{k}'_2|-\sigma}
          c^+_{\vec{k}-\vec{q}|-\sigma}c_{\vec{k}|-\sigma}\}
         \nonumber\\
 &+&\delta_{\sigma'|-\sigma}
      \{\delta_{\vec{k}_2|\vec{k}'-\vec{q}'}
          c^+_{\vec{k}_1\sigma}c_{\vec{k}'|\sigma}
          c^+_{\vec{k}'_1|-\sigma}c_{\vec{k}'_2|-\sigma}
          c^+_{\vec{k}-\vec{q}|-\sigma}c_{\vec{k}|-\sigma}
         \nonumber\\
 &&\qquad-\delta_{\vec{k}'_2|\vec{k}-\vec{q}}
          c^+_{\vec{k}'_1|-\sigma}c_{\vec{k}|-\sigma}
          c^+_{\vec{k}_1|\sigma}c_{\vec{k}_2|\sigma}
          c^+_{\vec{k}'-\vec{q}'|\sigma}c_{\vec{k}'|\sigma}\}\;.
\end{eqnarray}
The further evaluation of Eqs.\ (\ref{cf2interim}),
(\ref{cf3p2interim}) and (\ref{cf3p3interim}) requires the calculation
of expectation values. Fortunately, not all terms that arise from the 
corresponding factorizations, contribute: Terms proportional to 
$\delta_{\vec{q}|\vec{0}}$ or $\delta_{\vec{q'}|\vec{0}}$ may be
omitted in any case. And terms proportional to 
$\delta_{\vec{k}'_2|\vec{k}'_1-\vec{q'}}$ and
$\delta_{\vec{p}_1|\vec{p}_2}$ do not contribute in the case of
Eq.\ (\ref{cf2interim}) and Eq.\ (\ref{cf3p2interim}) along with Eq.\ 
(\ref{cf3p3interim}), respectively, due to the fact that $S(0)=0$.
Thus, we see for example that a factor
$f(\epsilon_{\vec{p}_1})(1-f(\epsilon_{\vec{p}_2}))$ may be split off
from the expectation value within the integrand of Eq.\ 
(\ref{cf3p2interim}). This factor combines with the quotient that
stems from the operator (\ref{fromSmatrix}) according to
\begin{equation} 
 \frac{e^{\beta(\epsilon_{\vec{p}_1}-\epsilon_{\vec{p}_2})}-1}
               {\epsilon_{\vec{p}_1}-\epsilon_{\vec{p}_2}}
  f(\epsilon_{\vec{p}_1})(1-f(\epsilon_{\vec{p}_2}))=
 -\frac{f(\epsilon_{\vec{p}_1})-f(\epsilon_{\vec{p}_2})}
       {\epsilon_{\vec{p}_1}-\epsilon_{\vec{p}_2}}\;.
\end{equation}
The further calculation is not simple and takes some time, especially
in the case of the functions (\ref{cf3p2interim}) and 
(\ref{cf3p3interim}). In the thermodynamic limit, where we may
replace, e.g., $N_y\delta_{k_y|0}\rightarrow2\pi\delta(k_y)$, 
$(1/N)\sum_{\vec{k}}\rightarrow\int_{1.\,BZ}d^dk/(2\pi)^d$ etc.,
our final result may be written in the form of Eqs.\ 
(\ref{RHz})-(\ref{LIII}). 

\section{Brillouin zone averages in infinite dimensions}

The calculation of Brillouin zone averages to leading order in
$1/d$ follows the procedure outlined in Ref.\ \cite{mueller}. 
In the following, Schl\"afli's integral representation of the Bessel
functions will play an important role: 
\begin{equation}
 <e^{ink+ir\cos k}>_k=i^{|n|}J_{|n|}(r)\equiv G_n(r)\;.
\label{schlaefli}
\end{equation}
Here, $n$ is an integer and we have used 
the notation $<\ldots>_k\equiv\int_{-\pi}^{\pi}\ldots\;dk/(2\pi)$. 
By means of the Fourier representation of the delta function, we may 
write, e.g., the function defined in Eq.\ (\ref{B_e}) as follows:
\begin{equation}
 B(\epsilon)=\int_{-\infty}^{\infty}\frac{ds}{2\pi}\;[iJ_1(r)]^2
             [J_0(r)]^{d-2}e^{is\epsilon}\;, 
\end{equation}
where here and in the following, $r\equiv s/\sqrt{d}$. Expanding the
Bessel functions in powers of $r$ and taking the limit
$d\rightarrow\infty$, we find:
\begin{equation}
 B(\epsilon)=\frac{1}{4d}\frac{\partial^2}{\partial\epsilon^2}
                                          D(\epsilon)\;.
\end{equation}
Thereby, we used $D(\epsilon)=\int_{-\infty}^{\infty}\frac{ds}{2\pi}\;
e^{-\frac{s^2}{4}+is\epsilon}$, which is derived analogously. This
proves Eq.\ (\ref{B_e}). The corresponding evaluation of Eqs.\ 
(\ref{LII}) and (\ref{LIII}) requires a Fourier series expansion 
of the cotangent:
\begin{equation}
 {\cal P}\cot(\frac{k}{2})=\frac{1}{i}\sum_{R\ne0}sgn(R)e^{ikR}\;\,.
\label{expansion1}
\end{equation}
Here, the sum is over all integers $R$, except for the zero. Since we
are working in the thermodynamic limit, $R$ may be considered to be 
a component of a lattice vector. To prove this representation, we 
start out with a known formula for the coefficients $b_n$ 
appearing in the ansatz
${\cal P}\cot(\frac{k}{2})=\sum_{n=1}^{\infty}b_n\sin(nk)$:
\begin{equation}
 \frac{b_n}{2}={\cal P}\int_{-\pi}^{\pi}\frac{dk}{2\pi}\;
               \cot(\frac{k}{2})\sin(nk).
\end{equation}
With the substitution $z\equiv e^{ik}$, we may perform the principal 
value integration by invoking the theorem of residues: The integration
contour
goes around the unit circle with the point $z=1$ being excluded. Thus,
we have to add the residue at $z=0$ to the half residue at $z=1$. We 
obtain $b_n=2$ for all positive integers $n$, which proves the
statement (\ref{expansion1}). In the following, we show how the Rhs.\ 
of Eq.\ (\ref{LIII}) is evaluated to leading order in $1/\sqrt{d}$. 
Writing the delta functions, that contain energies, in terms of 
Fourier integrals and introducing an additional momentum average 
$<(2\pi)^d\delta(\vec{k}+\vec{k}_2-\vec{k'}_1-\vec{k'}_2)\;
\ldots>_{\vec{k}}\;\equiv1$, the problem reduces to the calculation of 
the following expression:
\begin{eqnarray}
 \left.\right<
 &&e^{-i[s\epsilon_{\vec{k}}
        +s_1\epsilon_{\vec{k}_1}+s_2\epsilon_{\vec{k}_2}
        +s_1'\epsilon_{\vec{k}'_1}+s_2'\epsilon_{\vec{k}'_2}]
     }\;\nonumber\\
 &&\times 2\pi\delta(k_x+k_{2x}-k_{1x}'-k_{2x}')\;2\pi\delta(k_y-k_{1y})
        \prod_{j=3}^d 2\pi\delta(k_j+k_{2j}-k_{1j}'-k_{2j}')\nonumber\\
 &&\times\{\cos k_{1x}+\cos k_{2x}+\cos k'_{1x}+\cos k'_{2x}-
          [\sin k_{1x}+\sin k_{2x}-\sin k'_{1x}-\sin k'_{2x}]
        \,{\cal P}\cot(\frac{k_{1x}-k_{x}}{2})\}\nonumber\\
 &&\times\{\sin k_{1y}+\sin k_{2y}-\sin k'_{1y}-\sin k'_{2y}\}
      \sin k_{1y}2\pi\,\delta(k_{1y}+k_{2y}-k'_{1y}-k'_{2y})
 \left>_{\vec{k}\vec{k}_1\vec{k}_2\vec{k}'_1\vec{k}'_2\;.}\right.
\label{averageexample}
\end{eqnarray}
Since every dimension $j$ contributes a term 
$-(1/\sqrt{d})\cos k_j$ to the band dispersion $\epsilon(\vec{k})$, 
the expression (\ref{averageexample}) decomposes into $d$ factors. 
For example, one factor arises from all $x$-components:
\begin{eqnarray}
 \left.\right<
 &&e^{i[r\cos k_x
        +r_1\cos k_{1x}+r_2\cos k_{2x}
        +r_1'\cos k_{1x}'+r_2'\cos k_{2x}']
     }\;
     2\pi\delta(k_x+k_{2x}-k_{1x}'-k_{2x}')
     \nonumber\\
 &&\times\{\cos k_{1x}+\cos k_{2x}+\cos k'_{1x}+\cos k'_{2x}
     \nonumber\\
 &&-[\sin k_{1x}+\sin k_{2x}-\sin k'_{1x}-\sin k'_{2x}]
    \,{\cal P}\cot(\frac{k_{1x}-k_{x}}{2})\}
 \left>_{k_x k_{1x} k_{2x} k_{1x}' k_{2x}'\;.}\right.
\end{eqnarray}
This expression decomposes further into eight terms corresponding to 
the ones in the curled brackets. Each has to be evaluated with the 
Fourier series expansion (\ref{expansion1}) and that of the delta 
function:
\begin{equation}
 2\pi\delta(k)=\sum_{R=-\infty}^{\infty} e^{ikR}\;.
\end{equation}
For example, the term 
$-\sin k_{1x}\,{\cal P}\cot(\frac{k_{1x}-k_{x}}{2})$ gives rise to the
contribution 
\begin{equation}
 \sum_{n\ne0}\sum_m
 \mbox{sgn}(n)G_{m-n}(r)\frac{1}{2}[G_{n+1}(r_1)-G_{n-1}(r_1)]
   G_m(r_2)G_m(r_1')G_m(r_2')\;,
\end{equation}
where Eq.\ (\ref{schlaefli}) has been used. The leading order in
$1/\sqrt{d}$ reads:
\begin{equation}
 -G_1(r)G_0(r_1)G_0(r_2)G_0(r_1')G_0(r_2')\;.
\end{equation} 
This contribution to the expression (\ref{averageexample}) is of 
order $1/\sqrt{d}$, as are all the others. The subsequent Fourier 
integrals over the variables $s_i$ yield combinations of the 
functions (\ref{A_e}), (\ref{B_e}) and (\ref{DOS}) thus leading 
ultimately to Eq.\ (\ref{LIIIeva}). Eq.\ (\ref{LIIeva}) is proven 
analogously. 

\end{appendix}

%%%%%%%%%%%%%%%%%%%%%%%%%%%%%%%%%%%%%%%%%%%%%%%%%%
%%%%%%           FIGURE CAPTIONS
%%%%%%%%%%%%%%%%%%%%%%%%%%%%%%%%%%%%%%%%%%%%%%%%%%
\newpage
%%%%%%%%%%%
%%%%%%%%%%%

\begin{references}
\bibitem{pruschke2}A variety of transport properties was investigated
within the Hubbard model in $d=\infty$ including the Hall effect in
Ref.\ \cite{pruschke3}. These authors have estimated
$t^*\sim0.25\,\mbox{eV}$. However, their treatment of the Hall effect 
is based on a formula for the Hall conductivity that was derived by 
neglecting vertex corrections and which is restricted to
Lorentzian-shaped spectral functions only \cite{voruganti}. In
contrast to this, none of both problems arise in the case of the
ordinary conductivity. For instance, its vertex corrections vanish in 
infinite dimensions \cite{khurana}.
\end{references}
\end{document}